\documentclass[a4paper,12pt]{article}
\pdfoutput=1
\usepackage{fancyhdr}
\usepackage{multicol}
\usepackage{graphicx,subfigure,diagbox}
\usepackage{booktabs}
\usepackage{amssymb,bm,mathrsfs,bbm,amscd}
\usepackage[tbtags]{amsmath}
\usepackage{amsmath,tabu}
\usepackage{threeparttable}
\usepackage{lastpage}
\usepackage[numbers,sort&compress]{natbib}

\newlength{\dinwidth}
\newlength{\dinmargin}
\allowdisplaybreaks

\begin{document}

\fancyfoot[C]{\small \thepage}

\title{\vspace{-0.8cm}
\bf \boldmath{$K^0-\bar{K}^0$} mixing in the minimal flavor-violating two-Higgs-doublet models}

\author{%
      Natthawin Cho\footnote{natthawin.cho@mails.ccnu.edu.cn},
\quad Xin-Qiang Li\footnote{xqli@itp.ac.cn},
\quad Fang Su\footnote{sufang@itp.ac.cn},
\quad Xin Zhang\footnote{zhangxin027@mails.ccnu.edu.cn}\\
{\small Institute of Particle Physics and Key Laboratory of Quark and Lepton Physics~(MOE),}\\
{\small Central China Normal University, Wuhan, Hubei 430079, China}
}
\date{}
\maketitle

\begin{abstract}
\noindent The two-Higgs-doublet model (2HDM), as one of the simplest extensions of the Standard Model (SM), is obtained by adding another scalar doublet to the SM, and is featured by a pair of charged Higgs, which could affect many low-energy processes. In the ``Higgs basis" for a generic 2HDM, only one scalar doublet gets a nonzero vacuum expectation value and, under the criterion of minimal flavor violation, the other one is fixed to be either color-singlet or color-octet, which are named as the type-III and the type-C 2HDM,
respectively. In this paper, we study the charged-Higgs effects of these two models on the $K^0-\bar{K}^0$ mixing, an ideal process to probe New Physics (NP) beyond the SM. Firstly, we perform a complete one-loop computation of the box diagrams relevant to the $K^0-\bar{K}^0$ mixing, keeping the mass and momentum of the external strange quark up to the second order. Together with the up-to-date theoretical inputs, we then give a detailed phenomenological analysis, in the cases of both real and complex Yukawa couplings of the charged Higgs to quarks. The parameter spaces allowed by the current experimental data on the mass difference $\Delta m_K$ and the CP-violating parameter $\epsilon_K$ are obtained and the differences between these two 2HDMs are investigated, which are helpful to distinguish them from each other from a phenomenological point of view.
\end{abstract}

\section{Introduction}

The SM of particle physics has been proved to be successful because of its elegance and predictive capability. Almost all predictions in the SM are in good agreement with the experimental measurements, especially for the discovery of a Higgs boson with its mass around 125 GeV~\cite{Aad:2012tfa,Chatrchyan:2012xdj}. The discovery of a SM-like Higgs boson suggests that the electro-weak symmetry breaking (EWSB) is probably realized by the Higgs mechanism implemented via a single scalar doublet. However, the EWSB is not necessarily induced by just one scalar. It is interesting to note that many NP models are equipped with an extended scalar sector; for example, the minimal supersymmetric standard model requires at least two Higgs doublets~\cite{Carena:2002es}. Moreover, the SM does not provide enough sources of CP violation to generate the sufficient size of baryon asymmetry of the universe (BAU)~\cite{Trodden:1998qg,Dimopoulos:1978pw,Cline:2006ts}.

One of the simplest extensions of the SM scalar sector is the so-called 2HDM~\cite{Lee:1973iz}, in which a second scalar doublet is added to the SM field content. The added scalar doublet can provide additional sources of CP violation besides that from the Cabibbo-Kobayashi-Maskawa (CKM)~\cite{Cabibbo:1963yz, Kobayashi:1973fv} matrix, making it possible to explain the BAU~\cite{Trodden:1998qg}.

It is known that, within the SM, the flavour-changing neutral current (FCNC) interactions are forbidden at tree level, and are also highly suppressed at higher orders, due to the Glashow-Iliopoulos-Maiani
mechanism~\cite{Glashow:1970gm}. To avoid the experimental constraints on the FCNCs, the natural flavor conservation (NFC)~\cite{Glashow:1976nt} and minimal flavor violation (MFV)~\cite{Buras:2000dm,Buras:2010mh,DAmbrosio:2002vsn,Cervero:2012cx}  hypotheses have been proposed\footnote{The NFC and MFV hypotheses are not the only alternatives to avoid constraints from FCNCs; models with controlled FCNCs have also been addressed in the literature~\cite{Hadeed:1985xn,Branco:1996bq,Cvetic:1998uw,Bhattacharyya:2014nja,Campos:2017dgc}.}. In the NFC hypothesis, the absence of dangerous FCNCs is guaranteed by limiting the number of scalar doublets coupling to a given type of right-handed fermion to be at most one. This can be explicitly achieved by applying a discrete $\mathcal Z _2$ symmetry to the two scalar doublets differently, leading to four types of 2HDM (usually named as type-I, II, X and Y)~\cite{Branco:2011iw,Gunion:1989we}, which have been studied extensively for many years. In the MFV hypothesis, to control the flavor-violating interactions, all the scalar Yukawa couplings are assumed to be composed of the SM ones $Y^U$ and $Y^D$. In the ``Higgs basis"~\cite{Davidson:2005cw}, in which only one doublet gets a nonzero vacuum expectation value (VEV) and behaves the same as the SM one, the allowed $SU(3)\times SU(2)\times U(1)$ representation of the second scalar doublet is fixed to be either $(1,2)_{1/2}$ or $(8,2)_{1/2}$~\cite{Manohar:2006ga}, which implies that the second scalar doublet can be either color-singlet or color-octet. For convenience, they are referred as the type-III and the type-C model~\cite{Degrassi:2010ne}, respectively. Examples of the color-singlet case include the aligned 2HDM (A2HDM)~\cite{Pich:2009sp,Jung:2010ik} and the four types of 2HDM reviewed in Refs.~\cite{Branco:2011iw,Gunion:1989we}. In the color-octet case, the scalar spectrum contains one CP-even, color-singlet Higgs boson (the usual SM one) and three color-octet particles, one CP-even, one CP-odd and one electrically charged~\cite{Manohar:2006ga}.

Although the scalar-mediated flavor-violating interactions are protected by the MFV hypothesis, the type-III and type-C models can still bring in many interesting phenomena in some low-energy processes, especially due to the presence of a charged Higgs boson~\cite{Manohar:2006ga,Degrassi:2010ne,Li:2013vlx,Gresham:2007ri,Idilbi:2009cc,Bonciani:2007ex,Burgess:2009wm}. The neutral-meson mixings are of particular interest in this respect, because the charged-Higgs contributions to these processes arise at the same order as does the $W$ boson in the SM, indicating that the NP effects might be significant. For example, the charged-Higgs effects of these two models on the $B_s^0-\bar B_s^0$ mixing have been studied in Ref.~\cite{Li:2013vlx}. In this paper, we shall explore the $K^0-\bar K^0$ mixing within these two models and pursue possible differences between their effects. The general formula for $K^0-\bar K^0$ mixing, including the charged-Higgs contributions, could be found, for example, in Ref.~\cite{Crivellin:2013wna}.

Our paper is organized as follows. In Sec.~2, we review briefly the 2HDMs under the MFV hypothesis and give the theoretical framework for the $K^0-\bar K^0$ mixing. In Sec.~3, we perform a complete one-loop computation of the Wilson coefficients for the process within these two models. In Sec.~4, numerical results and discussions are presented in detail. Finally, our conclusions are made in Sec.~5. Explicit expressions for the loop functions appearing in the $K^0-\bar K^0$ mixing are collected in the appendix.

\section{Theoretical framework}
\label{sec:theoretical framework}

\subsection{Yukawa Sector}

Specifying to the ``Higgs basis"~\cite{Davidson:2005cw}, in which only one doublet gets a nonzero VEV, we can write the most general Lagrangian of Yukawa couplings between the two Higgs doublets, $\Phi_1$ and $\Phi_2$, and quarks as~\cite{Manohar:2006ga,Degrassi:2010ne}
\begin{equation}\label{Y}
-\mathcal{L}_Y=\bar{q}^0_L\tilde{\Phi}_1Y^Uu_R^0 +\bar{q}_L^0\Phi_1Y^Dd_R^0 +\bar{q}^0_L\tilde{\Phi}_2^{(a)}T_R^{(a)}\bar{Y}^Uu_R^0 +\bar{q}_L^0\Phi_2^{(a)}T_R^{(a)}\bar{Y}^Dd_R^0+\text{h.c.},
\end{equation}
where $\tilde{\Phi}_j=i\sigma_2\Phi^*_j~(j=1,2)$ with $\sigma_2$ the Pauli matrix, and $\bar{q}^0_L$, $u_R^0$, and $d_R^0$ are the quark fields given in the interaction basis. $T_R^{(a)}$ is the $SU(3)$ color generator which determines the color nature of the second Higgs doublet\footnote{Depending on which type of 2HDM we are considering, the second Higgs doublet can be either color-singlet or color-octet.}. $Y^{U,D}$ and $\bar{Y}^{U,D}$ are the Yukawa couplings and are generally complex $3\times3$ matrices in the quark flavor space.

According to the MFV hypothesis, the transformation properties of the Yukawa coupling matrices $Y^{U,D}$ and $\bar{Y}^{U,D}$ under the quark flavor symmetry group $SU(3)_{Q_L}\otimes SU(3)_{U_R}\otimes SU(3)_{D_R}$ are required to be the same. This can be achieved by requiring $\bar Y^{U,D}$ to be composed of pairs of $Y^{U,D}$~\cite{Degrassi:2010ne}
\begin{equation}
\begin{aligned}
\bar{Y}^U&=A_u^*(1+\epsilon_u^*Y^UY^{U\dagger}+...)Y^U,\\
\bar{Y}^D&=A_d(1+\epsilon_dY^UY^{U\dagger}+...)Y^D.
\end{aligned}
\end{equation}

Transforming the Lagrangian in Eq.~\eqref{Y} from the interaction basis to the mass basis, one can obtain the Yukawa interactions of charged Higgs with quarks in the mass-eigenstate basis, which are given by~\cite{Degrassi:2010ne,Li:2013vlx}
\begin{equation}
\mathcal{L}_{H^+}=\dfrac{g}{\sqrt{2}m_W}\sum_{i,j=1}^{3}\bar{u}_iT_R^{(a)}(A^i_um_{u_i}P_L-A^i_dm_{d_j}P_R)V_{ij}d_jH^+_{(a)}+\text{h.c.},
\end{equation}
where $A^i_{u,d}$ are family-dependent Yukawa coupling constants~\cite{Degrassi:2010ne,Li:2013vlx,Chang:2015rva}
\begin{equation}
A^i_{u,d}=A_{u,d}\left(1+\epsilon_{u,d}\dfrac{m_t^2}{v^2}\delta_{i3}\right),
\end{equation}
with $v=\langle\Phi_1^0\rangle=174$~GeV. For simplicity, we consider only the family universal coupling case in which the family-dependent Yukawa couplings, $A^i_{u,d}$, can be simplified to $A^i_{u,d}=A_{u,d}$.

\subsection{$K^0-\bar{K}^0$ mixing}

\begin{figure}[t]
\begin{center}
  \subfigure[]{\includegraphics[width=1.6in]{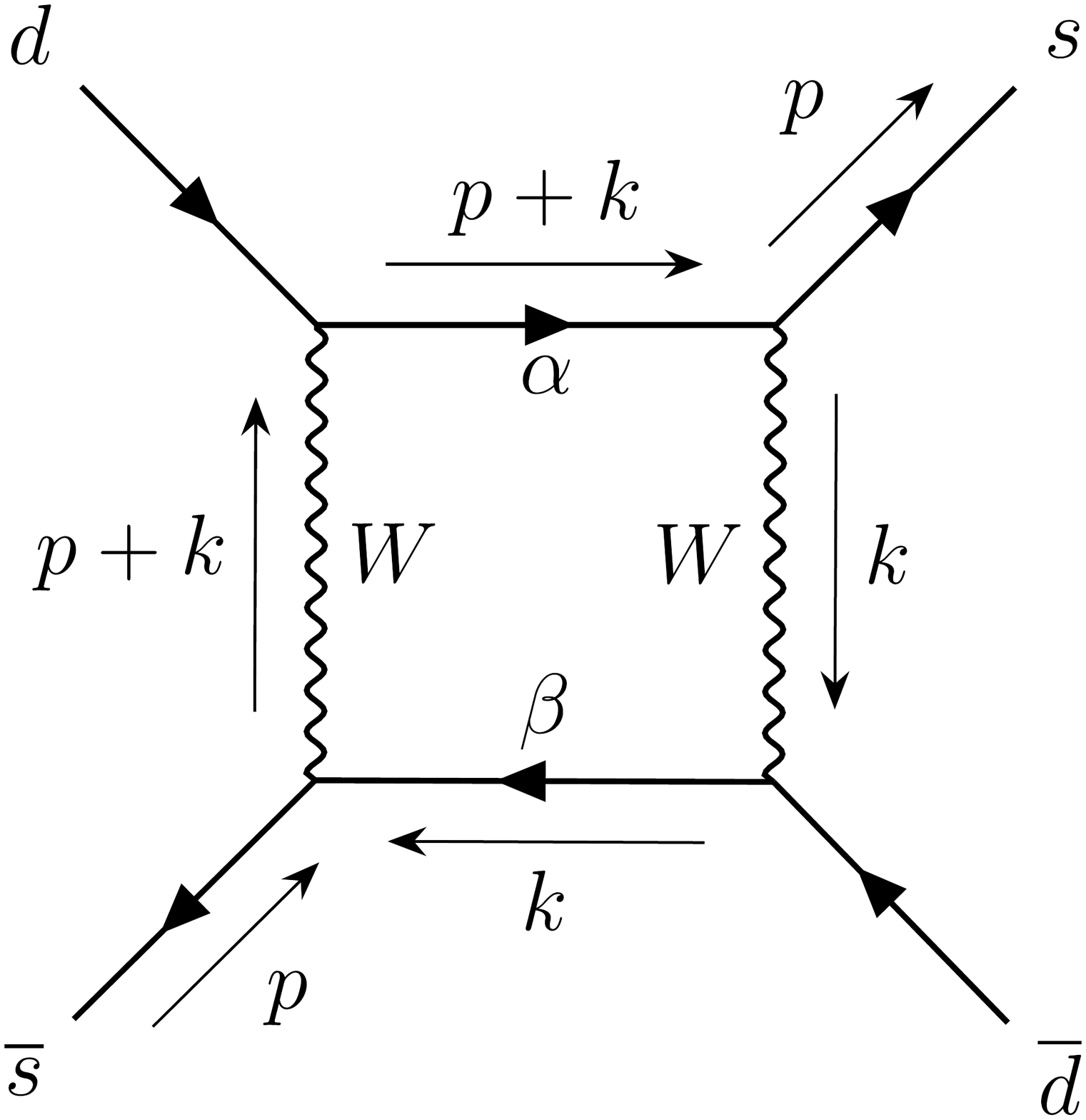}}~
  \subfigure[]{\includegraphics[width=1.6in]{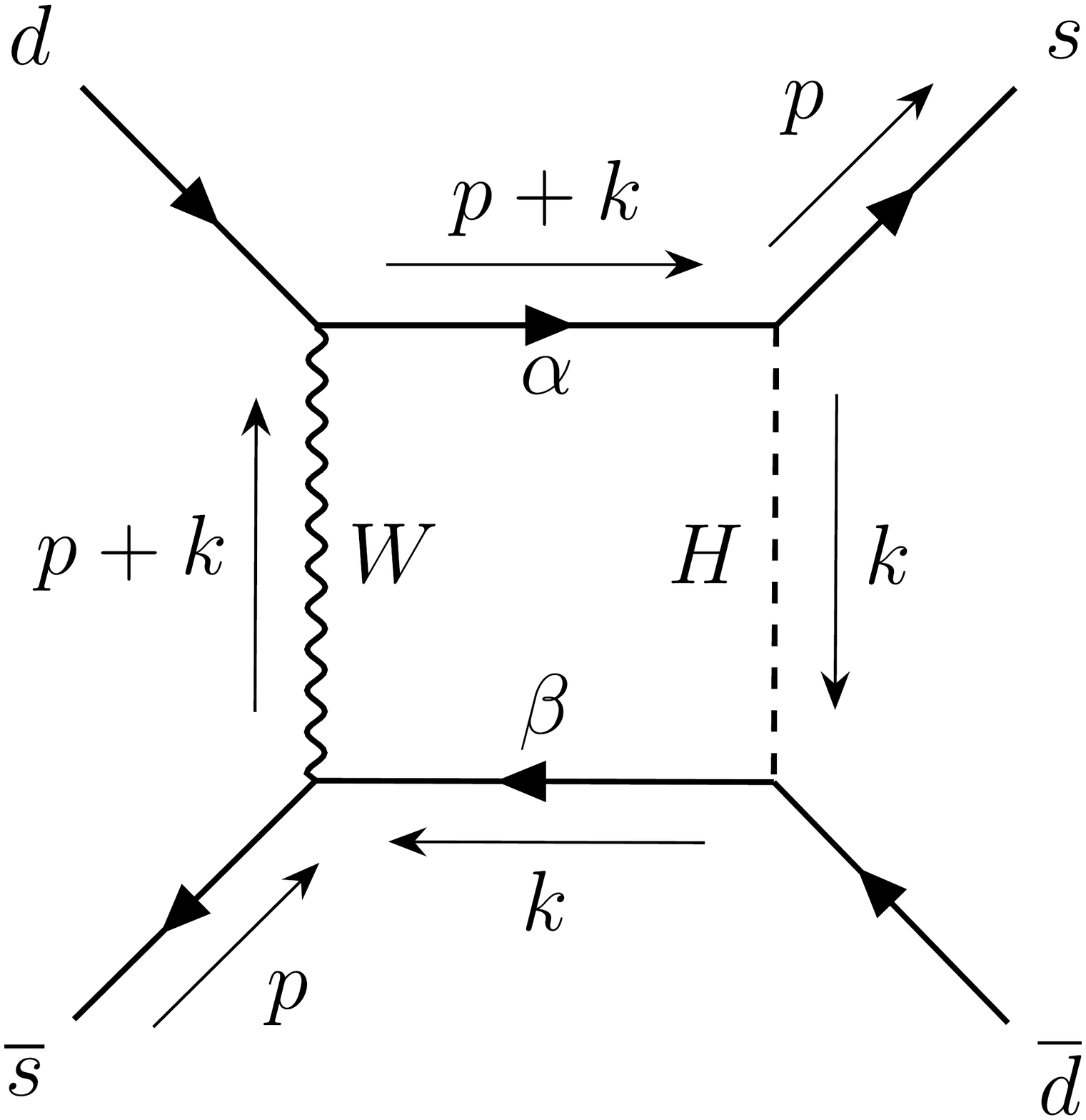}}~
  \subfigure[]{\includegraphics[width=1.6in]{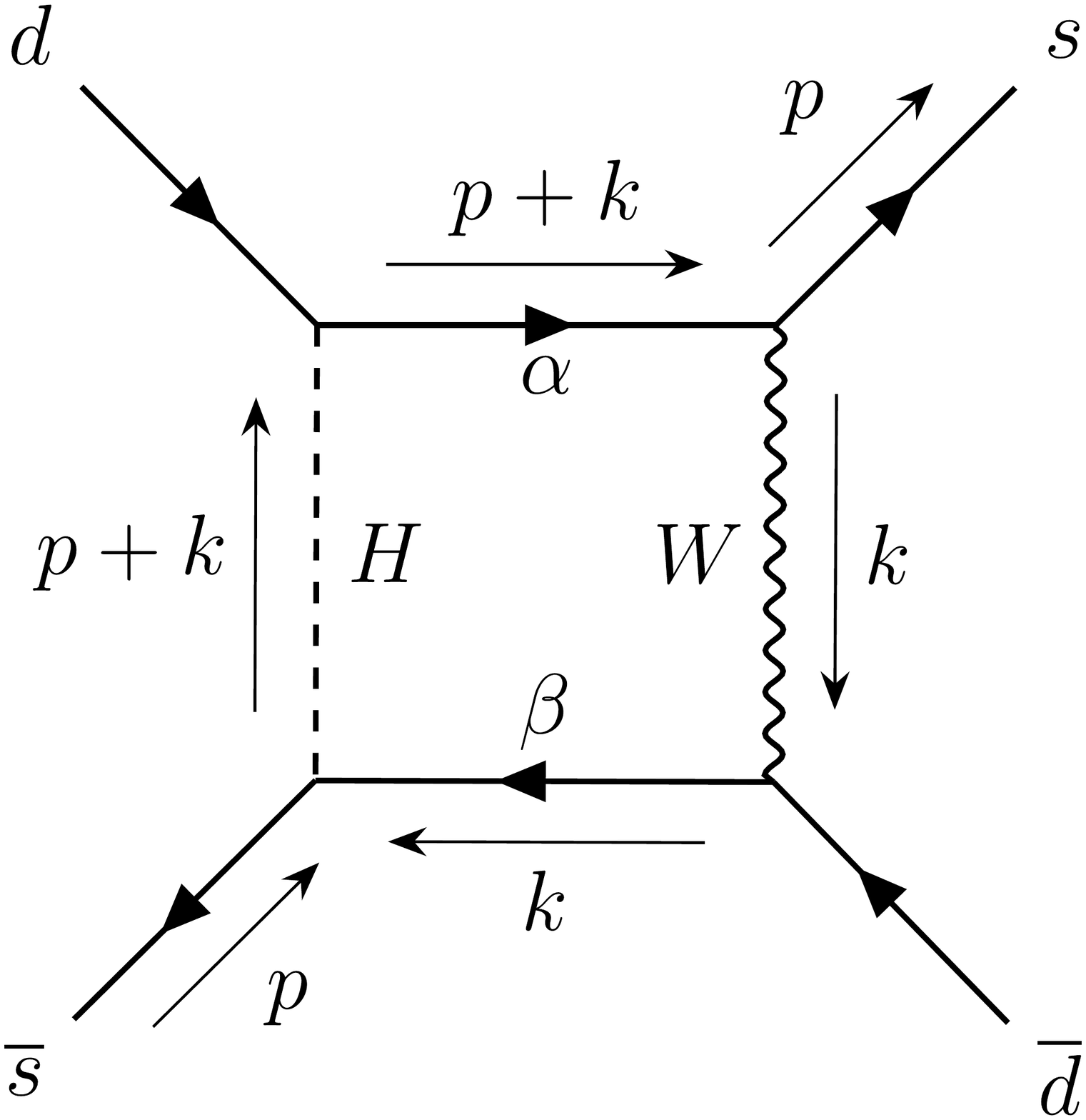}}~
  \subfigure[]{\includegraphics[width=1.6in]{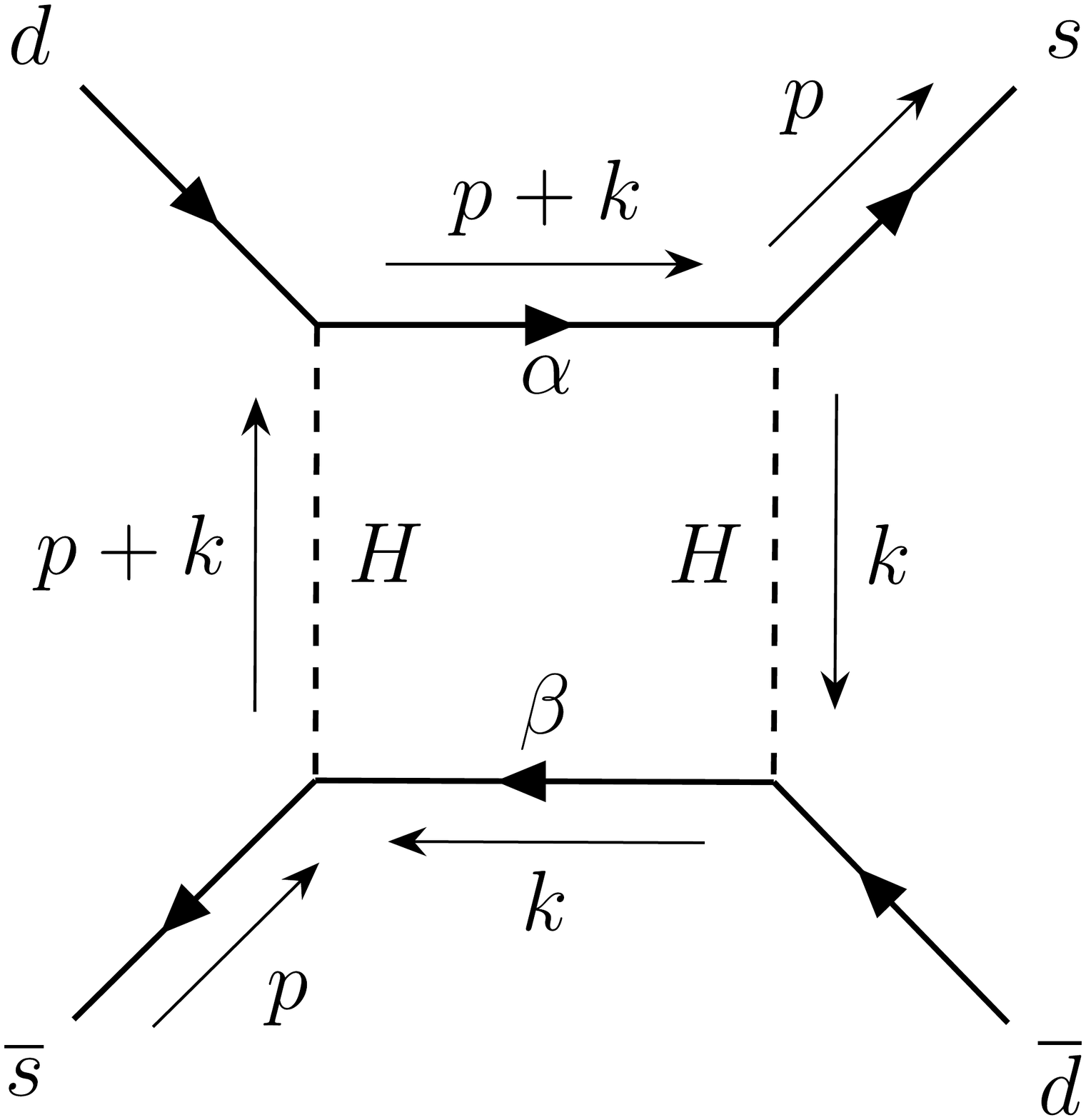}}
  \caption{\label{box} \small Box diagrams for the $K^0-\bar{K}^0$ mixing in the unitary gauge both within the SM ((a)) and in the 2HDMs with MFV ((b)-(d)). Crossed diagrams, which are related to the original ones by interchanging the external lines, have also been taken into account.}
\end{center}
\end{figure}

Both within the SM and in the 2HDMs with MFV, the neutral kaon mixing occurs via the box diagrams depicted in Fig.~\ref{box}\footnote{These Feynman diagrams are drawn with the LaTeX package \emph{TikZ-Feynman}~\cite{Ellis:2016jkw}.}. As demonstrated in Ref.~\cite{Urban:1997cm}, the correction from the external momenta and quark masses is not negligible for the $K^0-\bar{K}^0$ mixing. Thus, unlike the traditional calculation performed in the limit of vanishing external momenta and external quark masses, we shall keep the external strange-quark momentum and mass to the second order; this is essential to guarantee the final result gauge-independent~\cite{Chang:2015rva}.

Calculating the one-loop box diagrams and following the standard procedure of matching~\cite{Urban:1997cm}, we obtain the effective Hamiltonian responsible for the $K^0-\bar{K}^0$ mixing
\begin{equation}\label{eqn:Hamiltonian}
\mathcal{H}_\text{eff}=\dfrac{G_F^2m_W^2}{16\pi^2}\Big[C^{VLL}(\mu)Q^{VLL} +C^{SLL}(\mu)Q^{SLL}+C^{TLL}(\mu)Q^{TLL}\Big]+\text{h.c.},
\end{equation}
where $G_F$ is the Fermi coupling constant, $m_W$ the $W$-boson mass, and $C_i(\mu)$ the scale-dependent Wilson coefficients of the four-quark operators $Q_i$, which are defined, respectively, as\footnote{There are totally eight four-quark operators for the most general case~\cite{Buras:2001ra}, but we have written out only the operators that exist in our calculation.}
\begin{equation}
\begin{aligned}
Q^{VLL}&=\bar{s}^\alpha\gamma_\mu(1-\gamma_5)d^\alpha\,\bar{s}^\beta\gamma^\mu(1-\gamma_5)d^\beta,\\
Q^{SLL}&=\bar{s}^\alpha(1-\gamma_5)d^\alpha\,\bar{s}^\beta(1-\gamma_5)d^\beta,\\
Q^{TLL}&=\bar{s}^\alpha\sigma_{\mu\nu}(1-\gamma_5)d^\alpha\,\bar{s}^\beta\sigma^{\mu\nu}(1-\gamma_5)d^\beta,
\end{aligned}
\end{equation}
with $\alpha$ and $\beta$ the color indices and $\sigma_{\mu\nu}=\frac{1}{2}[\gamma_\mu,\gamma_\nu]$. Note that we include the QCD corrections only to the SM Wilson coefficient $C^{VLL}$, but not to the NP ones. The hadronic matrix elements of these operators can be written as~\cite{Buras:2001ra}
\begin{align}
\langle\bar{K}^0|Q^{VLL}|K^0\rangle&=\dfrac{4}{3}\,m_KF_K^2B_1^{VLL}(\mu),\\
\langle\bar{K}^0|Q^{SLL}|K^0\rangle&=-\dfrac{5}{6}\,R(\mu)m_KF_K^2B_1^{SLL}(\mu),\\
\langle\bar{K}^0|Q^{TLL}|K^0\rangle&=-2\,R(\mu)m_KF_K^2B_2^{SLL}(\mu),
\end{align}
where $m_K$ is the kaon mass, and $F_K$ the kaon decay constant. $B_i^j(\mu)$ is the scale-dependent bag parameters, and $R(\mu)$ is defined as~\cite{Buras:2001ra}
\begin{equation}\label{R}
R(\mu)=\left(\dfrac{m_K}{m_s(\mu)+m_d(\mu)}\right)^2.
\end{equation}

It should be noted that the SM and NP contributions to the Wilson coefficients $C^i(\mu)$ cannot be summed directly because they are given at different initial scales, $\mu_W$ for the SM and $\mu_{H^{\pm}}$ for the 2HDM in particular. In order to sum these two contributions, they must be firstly run down to the lattice scale at which the bag parameters $B_i^j(\mu)$ are evaluated. The explicit expressions for these Wilson coefficients will be presented in Sec.~3.

For the $K^0-\bar{K}^0$ mixing, there exist two observables which can be calculated from the effective Hamiltonian given by Eq.~\eqref{eqn:Hamiltonian}~\cite{Buras:2001ra}
\begin{align}\label{mk}
\Delta m_K &=2\,\text{Re}\langle K^0|\mathcal{H}_{\text{eff}}|\bar{K}^0\rangle,\\[0.2cm]
\label{ek}
\epsilon_K &=\frac{\exp(i\pi/4)}{\sqrt{2}\Delta m_K}\,\text{Im}\langle K^0|\mathcal{H}_{\text{eff}}|\bar{K}^0\rangle.
\end{align}

The above equations are the most general formulae for these two observables. It should be noted that $\Delta m_K$ and $\epsilon_K$ receive both short-distance (SD) and long-distance (LD) contributions. With the LD contribution included, the mass difference $\Delta m_K$ can be decomposed as~\cite{Buras:2010pza}
\begin{equation}
\Delta m_K=\Delta m_K^{SD}+\Delta m_K^{LD}|_{\pi\pi}+\Delta m_K^{LD}|_{\eta'},
\end{equation}
where the SD part is derived from Eq.~\eqref{mk} with the effective Hamiltonian obtained from the box diagrams, while the two LD parts are estimated, respectively, as~\cite{Buras:2010pza,Gerard:2005yk}
\begin{equation}\label{mkld}
\Delta m_K^{LD}|_{\pi\pi}=0.4\,\Delta m_K^{\rm exp}, \qquad \Delta m_K^{LD}|_{\eta'}=-0.3\,\Delta m_K^{\rm exp}.
\end{equation}

We can see from Eq.~\eqref{mkld} that the LD contribution to $\Delta m_K$ is about $10\%$ of the experimental value. However, keeping in mind that this estimate is just a bold-guess based on an analysis at the leading chiral logarithm in the framework of chiral perturbation theory~\cite{Buras:2010pza}, we should note that the actual uncertainty on $\Delta m_K^{LD}$ is quite huge\footnote{We thank Prof. Antonio Pich for pointing out this to us.}. As the structure of LD contribution is still not well understood, we include this part only in the SM case but not in the NP one.

The formula for the CP-violating parameter $\epsilon_K$, with the LD contribution taken into account, is given by~\cite{Ligeti:2016qpi}
\begin{equation}\label{ek_LD}
\epsilon_K=\dfrac{\kappa_\epsilon e^{i\phi_\epsilon}}{\sqrt{2}} \dfrac{\text{Im}M_{12}^{SD}}{\Delta m_K^{exp}},
\end{equation}
where $\kappa_\epsilon=0.94~(2)$~\cite{Buras:2010pza}, $\phi_\epsilon=43.52~(5)^\circ$~\cite{Aoki:2016frl}, and $M_{12}^{SD}=\langle K^0|\mathcal{H}_{\text{eff}}|\bar{K}^0\rangle$. The LD contribution to $\epsilon_K$ has been included in the two phenomenological factors $\kappa_\epsilon$ and $\phi_\epsilon$. In the case with only the SD contribution, $\kappa_\epsilon=1$ and $\phi_\epsilon=\pi/4$, and Eq.~\eqref{ek_LD} goes back to Eq.~\eqref{ek}.

\section{Analytic calculation}
\label{sec:analytic calculation}

\subsection{Wilson coefficients within the SM}

For the SM case, we calculate the Wilson coefficients from the box diagram shown in Fig.~\ref{box}(a). Without any QCD correction, they are given, respectively, as
\begin{equation}\label{SMVLL}
\begin{split}
C^{VLL}_{SM}(\mu_W)&=\left[\lambda_c^2S_0(x_c)+\lambda_t^2S_0(x_t)+2\lambda_c\lambda_tS_0(x_c,x_t)\right]\\
&+x_s\left[\lambda_c^2f_1(x_c)+\lambda_t^2f_1(x_t)+2\lambda_c\lambda_tf_1(x_c,x_t)\right],
\end{split}
\end{equation}
\begin{equation}\label{SMSLL}
C^{SLL}_{SM}(\mu_W)=x_s\left[\lambda_c^2f_2(x_c)+\lambda_t^2f_2(x_t)+2\lambda_c\lambda_tf_2(x_c,x_t)\right],
\end{equation}
\begin{equation}\label{SMTLL}
C^{TLL}_{SM}(\mu_W)=x_s\left[\lambda_c^2f_3(x_c)+\lambda_t^2f_3(x_t)+2\lambda_c\lambda_tf_3(x_c,x_t)\right],
\end{equation}
where $x_i=\frac{m_i^2(\mu)}{m_W^2}$, and $S_0$ is the Inami-Lim function given by Eq.~\eqref{S0}~\cite{Inami:1980fz}. Explicit expressions for the functions $f_i$ can be found in the appendix. Note that, when the external strange-quark momentum and mass are kept to the second order, we also get nonzero contributions to the Wilson coefficients $C^{SLL}_{SM}$ and $C^{TLL}_{SM}$ even in the SM case.

The QCD corrections to the Wilson coefficients can be described by the factors $\eta_{cc}$, $\eta_{ct}$ and $\eta_{tt}$, which have been calculated up to the next-to-next-to-leading order~\cite{Buras:1990fn,Brod:2010mj,Brod:2011ty} and are collected in Ref.~\cite{Bobeth:2016llm}. Combining the renormalization group (RG) evolution with these QCD corrections, we get
\begin{align}
\langle\bar{K}^0|\mathcal{H}_{\text{eff}}|K^0\rangle^{VLL}_{SM} &=\zeta\bigg[\hat{B}\Big(\lambda_c^2\eta_{cc}S_0(\bar{x_c}) +\lambda_t^2\eta_{tt}S_0(\bar{x_t})+2\lambda_c\lambda_t\eta_{ct} S_0(\bar{x_c},\bar{x_t})\Big)\nonumber\\
&+P_{SM}^{VLL}x_{s,\mu_W}\Big(\lambda_c^2f_1(x_{c,\mu_W}) +\lambda_t^2f_1(x_{t,\mu_W})+2\lambda_c\lambda_t f_1(x_{c,\mu_W},x_{t,\mu_W})\Big)\bigg],\\[0.2cm]
\langle\bar{K}^0|\mathcal{H}_{\text{eff}}|K^0\rangle^{SLL}_{SM} &=\zeta\bigg[P_{SM}^{SLL}C^{SLL}_{SM}(\mu_W)\bigg],\\[0.2cm]
\langle\bar{K}^0|\mathcal{H}_{\text{eff}}|K^0\rangle^{TLL}_{SM}&=\zeta\bigg[P_{SM}^{TLL}C^{TLL}_{SM}(\mu_W)\bigg],
\end{align}
where $\zeta=\frac{G_F^2 m_W^2m_KF_K^2}{12\pi^2}$, and $\bar{x_i}\equiv\left(\frac{m_i(m_i)}{m_W}\right)^2$ is the scale-independent mass ratio, whereas $x_{i,\mu}\equiv\left(\frac{m_i(\mu)}{m_W}\right)^2$ is the mass ratio at the scale $\mu$. $\hat{B}$ is the RG independent bag parameter, and the factors $P^i$ encode the RG evolution effects that are given, respectively, as~\cite{Buras:2001ra}
\begin{align}\label{PVLLSM}
P^{VLL}_{SM}&=\Big[\eta(3\text{ GeV})\Big]_{VLL}^{SM}B_1(3\text{ GeV})\,,\\[0.2cm]
\label{PSLLSM}
P^{SLL}_{SM}&=-\frac{5}{8}\Big[\eta_{11}(3\text{ GeV})\Big]_{SLL}^{SM}\Big[B_2(3\text{ GeV})\Big]_{\text{eff}}-\frac{3}{2}\Big[\eta_{21}(3\text{ GeV})\Big]_{SLL}^{SM}\Big[B_3(3\text{ GeV})\Big]_{\text{eff}}\,,\\[0.2cm]
\label{PTLLSM}
P^{TLL}_{SM} &=-\frac{5}{8}\Big[\eta_{12}(3\text{ GeV})\Big]_{SLL}^{SM}\Big[B_2(3\text{ GeV})\Big]_{\text{eff}}-\frac{3}{2}\Big[\eta_{22}(3\text{ GeV})\Big]_{SLL}^{SM}\Big[B_3(3\text{ GeV})\Big]_{\text{eff}}\,,
\end{align}
where the effective bag parameters $\Big[B_i(3\text{ GeV})\Big]_{\text{eff}}$ are defined as~\cite{Buras:2001ra}
\begin{equation}
\Big[B_i(3\text{ GeV})\Big]_{\text{eff}}\equiv R(3\text{ GeV})B_i(3\text{ GeV}),
\end{equation}
with $R(\mu)$ defined in Eq.~\eqref{R}. The factors $\eta$ and $\eta_{i,j}$ are given by the formulae collected in Ref.~\cite{Buras:2001ra} with
\begin{equation}
\eta_4\equiv\dfrac{\alpha_s^{(4)}(\mu_b)}{\alpha_s^{(4)}(3\text{ GeV})}, \quad \quad \eta_5\equiv\dfrac{\alpha_s^{(5)}(\mu_W)}{\alpha_s^{(5)}(\mu_b)}.
\end{equation}

\subsection{Wilson coefficients in the 2HDMs with MFV}

The Wilson coefficients at the matching scale $\mu_{H^\pm}\sim m_{H^\pm}$ in the NP case are calculated from the box diagrams shown in Figs.~\ref{box}(b)--\ref{box}(d), with the results given, respectively, as
\begin{align}
\begin{split}
C_{III}^{VLL}(\mu_{H^\pm})&=A_uA_u^*\bigg[\lambda_c^2\Big(f_4(x_c,x_H)+x_sg_4(x_c,x_H)\Big)+\lambda_t^2\Big(f_4(x_t,x_H)+x_sg_4(x_t,x_H)\Big)\\
&\hspace{0.7cm}+2\lambda_c\lambda_t\Big(f_4(x_c,x_t,x_H)+x_sg_4(x_c,x_t,x_H)\Big)\bigg]\\
&\hspace{-0.35cm}+A_uA_d^*x_s\bigg[\lambda_c^2f_5(x_c,x_H)+\lambda_t^2f_5(x_t,x_H)+2\lambda_c\lambda_tf_5(x_c,x_t,x_H)\bigg]\\
&\hspace{-0.5cm}+A_u^2A_u^{*2}x_s\bigg[\lambda_c^2f_6(x_c,x_H)+\lambda_t^2f_6(x_t,x_H)+2\lambda_c\lambda_tf_6(x_c,x_t,x_H)\bigg],
\end{split}\\[0.2cm]
C_{III}^{SLL}(\mu_{H^\pm})&=x_s\bigg[A_uA_u^*\Big(\lambda_c^2\big[f_7(x_c,x_H)+g_7(x_c,x_H)\big]+\lambda_t^2\big[f_7(x_t,x_H)+g_7(x_t,x_H)\big]\nonumber\\
&\hspace{1.4cm}+2\lambda_c\lambda_t\big[f_7(x_c,x_t,x_H)+g_7(x_c,x_t,x_H)\big]\Big)\nonumber\\
&\hspace{0.7cm}+A_uA_d^*\Big(\lambda_c^2f_8(x_c,x_H)+\lambda_t^2f_8(x_t,x_H)+2\lambda_c\lambda_tf_8(x_c,x_t,x_H)\Big)\nonumber\\
&\hspace{0.6cm}+A_u^2A_u^{*2}\Big(\lambda_c^2f_9(x_c,x_H)+\lambda_t^2f_9(x_t,x_H)+2\lambda_c\lambda_tf_9(x_c,x_t,x_H)\Big)\nonumber\\
&\hspace{0.55cm}+A_u^2A_d^{*2}\Big(\lambda_c^2f_{10}(x_c,x_H)+\lambda_t^2f_{10}(x_t,x_H)+2\lambda_c\lambda_tf_{10}(x_c,x_t,x_H)\Big)\nonumber\\
&\hspace{0.2cm}-A_u^2A_u^*A_d^*\Big(\lambda_c^2f_{10}(x_c,x_H)+\lambda_t^2f_{10}(x_t,x_H)+2\lambda_c\lambda_tf_{10}(x_c,x_t,x_H)\Big)\bigg],\\[0.2cm]
C_{III}^{TLL}(\mu_{H^\pm})&=0,\\[0.2cm]
\begin{split}
C_{C}^{VLL}(\mu_{H^\pm})&=\frac{1}{3}A_uA_u^*\bigg[\lambda_c^2\Big(f_4(x_c,x_H)+x_sg_4(x_c,x_H)\Big)+\lambda_t^2\Big(f_4(x_t,x_H)+x_sg_4(x_t,x_H)\Big)\\
&\hspace{1.0cm}+2\lambda_c\lambda_t\Big(f_4(x_c,x_t,x_H)+x_sg_4(x_c,x_t,x_H)\Big)\bigg]\\
&\hspace{-0.3cm}+\frac{1}{3}A_uA_d^*x_s\bigg[\lambda_c^2f_5(x_c,x_H)+\lambda_t^2f_5(x_t,x_H)+2\lambda_c\lambda_tf_5(x_c,x_t,x_H)\bigg]\\
&\hspace{-0.65cm}+\frac{11}{18}A_u^2A_u^{*2}x_s\bigg[\lambda_c^2f_6(x_c,x_H)+\lambda_t^2f_6(x_t,x_H)+2\lambda_c\lambda_tf_6(x_c,x_t,x_H)\bigg],
\end{split}\\[0.2cm]
\begin{split}
C_{C}^{SLL}(\mu_{H^\pm})&=x_s\bigg[-\frac{5}{12}A_uA_u^*\Big(\lambda_c^2\big[f_7(x_c,x_H)+g_7(x_c,x_H)\big]+\lambda_t^2\big[f_7(x_t,x_H)+g_7(x_t,x_H)\big]\\
&\hspace{2.4cm}+2\lambda_c\lambda_t\big[f_7(x_c,x_t,x_H)+g_7(x_c,x_t,x_H)\big]\Big)\\
&\hspace{1.2cm}-\frac{5}{12}A_uA_d^*\Big(\lambda_c^2f_8(x_c,x_H)+\lambda_t^2f_8(x_t,x_H)+2\lambda_c\lambda_tf_8(x_c,x_t,x_H)\Big)\\
&\hspace{1.1cm}-\frac{19}{72}A_u^2A_u^{*2}\Big(\lambda_c^2f_9(x_c,x_H)+\lambda_t^2f_9(x_t,x_H)+2\lambda_c\lambda_tf_9(x_c,x_t,x_H)\Big)\\
&\hspace{1.05cm}-\frac{19}{72}A_u^2A_d^{*2}\Big(\lambda_c^2f_{10}(x_c,x_H)+\lambda_t^2f_{10}(x_t,x_H)+2\lambda_c\lambda_tf_{10}(x_c,x_t,x_H)\Big)\\
&\hspace{0.7cm}+\frac{19}{72}A_u^2A_u^*A_d^*\Big(\lambda_c^2f_{10}(x_c,x_H)+\lambda_t^2f_{10}(x_t,x_H)+2\lambda_c\lambda_tf_{10}(x_c,x_t,x_H)\Big)\bigg],
\end{split}\\[0.2cm]
\begin{split}
C_{C}^{TLL}(\mu_{H^\pm})&=x_s\bigg[\frac{1}{16}A_uA_u^*\Big(\lambda_c^2\big[f_7(x_c,x_H)+g_7(x_c,x_H)\big]+\lambda_t^2\big[f_7(x_t,x_H)+g_7(x_t,x_H)\big]\\
&\hspace{1.9cm}+2\lambda_c\lambda_t\big[f_7(x_c,x_t,x_H)+g_7(x_c,x_t,x_H)\big]\Big)\\
&\hspace{0.7cm}+\frac{1}{16}A_uA_d^*\Big(\lambda_c^2f_8(x_c,x_H)+\lambda_t^2f_8(x_t,x_H)+2\lambda_c\lambda_tf_8(x_c,x_t,x_H)\Big)\\
&\hspace{0.6cm}+\frac{7}{96}A_u^2A_u^{*2}\Big(\lambda_c^2f_9(x_c,x_H)+\lambda_t^2f_9(x_t,x_H)+2\lambda_c\lambda_tf_9(x_c,x_t,x_H)\Big)\\
&\hspace{0.55cm}+\frac{7}{96}A_u^2A_d^{*2}\Big(\lambda_c^2f_{10}(x_c,x_H)+\lambda_t^2f_{10}(x_t,x_H)+2\lambda_c\lambda_tf_{10}(x_c,x_t,x_H)\Big)\\
&\hspace{0.2cm}-\frac{7}{96}A_u^2A_u^*A_d^*\Big(\lambda_c^2f_{10}(x_c,x_H)+\lambda_t^2f_{10}(x_t,x_H)+2\lambda_c\lambda_tf_{10}(x_c,x_t,x_H)\Big)\bigg].
\end{split}
\end{align}

Explicit expressions for the functions $f_i$ introduced in the above equations are collected in the appendix. Note that the contribution to $C^{TLL}$ is zero for the type-III but is not for the type-C 2HDM. With the RG evolution effect included, the final result is similar to the SM case and can be written as
\begin{align}
\langle\bar{K}^0|\mathcal{H}_{\text{eff}}|K^0\rangle^{VLL}_{III}&=\zeta\bigg[P^{VLL}_{NP}C^{VLL}_{III}(\mu_t)\bigg],\\[0.2cm]
\langle\bar{K}^0|\mathcal{H}_{\text{eff}}|K^0\rangle^{SLL}_{III}&=\zeta\bigg[P^{SLL}_{NP}C^{SLL}_{III}(\mu_t)\bigg],\\[0.2cm]
\langle\bar{K}^0|\mathcal{H}_{\text{eff}}|K^0\rangle^{TLL}_{III}&=0,
\end{align}
for the type-III, and
\begin{align}
\langle\bar{K}^0|\mathcal{H}_{\text{eff}}|K^0\rangle^{VLL}_{C}&=\zeta\bigg[P^{VLL}_{NP}C^{VLL}_{C}(\mu_t)\bigg],\\[0.2cm]
\langle\bar{K}^0|\mathcal{H}_{\text{eff}}|K^0\rangle^{SLL}_{C}&=\zeta\bigg[P_{NP}^{SLL}C^{SLL}_{C}(\mu_t)\bigg],\\[0.2cm]
\langle\bar{K}^0|\mathcal{H}_{\text{eff}}|K^0\rangle^{TLL}_{C}&=\zeta\bigg[P^{TLL}_{NP}C^{TLL}_{C}(\mu_t)\bigg],
\end{align}
for the type-C 2HDM. The factors $P^i$ are also similar to the SM case but with a different factor $\eta_5$, which is now defined by
\begin{equation}
\eta_5\equiv\dfrac{\alpha_s^{(5)}(\mu_t)}{\alpha_s^{(5)}(\mu_b)}.
\end{equation}
Here the matching scale for the 2HDMs has been changed to $\mu_t\sim m_t$, because the evolution effect from $\mu_{H^\pm}\sim m_{H^\pm}$ down to $\mu_t\sim m_t$ is quite small and can be safely neglected.

After performing the proper RG evolution, we can then sum directly both the SM and NP contributions to the matrix element $\langle\bar{K}^0|\mathcal{H}_{\text{eff}}|K^0\rangle$, which can be written as
\begin{equation}\label{sumWC}
\begin{split}
\langle\bar{K}^0|\mathcal{H}_{\text{eff}}|K^0\rangle^{i}=\langle\bar{K}^0|
\mathcal{H}_{\text{eff}}|K^0\rangle^{i}_{SM}+\langle\bar{K}^0|\mathcal{H}_{\text{eff}}|K^0\rangle^{i}_{NP}\,,
\end{split}
\end{equation}
where the superscript `$i$' labels the different four-quark operators.

\section{Numerical results and discussions}
\label{sec:numerical analysis}

\subsection{Input parameters and the SM results}

\begin{table}[htbp]
\begin{center}
\caption{\label{table:input} \small Input parameters used throughout this paper, together with the experimental data.}
\vspace{0.2cm}
\tabulinesep=0.5mm
\begin{tabu} to\linewidth {|X[2l] X[2l]|}
\hline\hline
\multicolumn{2}{|c|}{Electro-weak parameters~\cite{Olive:2016xmw}} \\
$m_W$ = 80.385(15) GeV & $m_Z$ = 91.1876(21) GeV \\
$\mu_W$ = 80.385 GeV& $\mu_t$ = 163.427 GeV $^a$\\
$G_F$ = 1.1663787(6) $\times 10^{-5}$ GeV$^{-2}$ &  \\
\end{tabu}
\begin{tabu} to\linewidth {|X[2l] X[2l]|}
\hline
\multicolumn{2}{|c|}{QCD coupling constant} \\
$\alpha_s(\mu_t)$ = 0.1086(10) $^a$ & $\alpha_s(m_Z)$ = 0.1182(12)~\cite{Olive:2016xmw} \\
$\alpha_s(\mu_W)$ = 0.1205(12) $^a$ & $\alpha_s(\mu_b)$ = 0.2243(45) $^a$ \\
$\alpha_s($3 GeV$)$ = 0.2521$^{+0.0058}_{-0.0057}$ $^a$ &  \\
\end{tabu}
\begin{tabu} to\linewidth {|X[2l] X[2l]|}
\hline
\multicolumn{2}{|c|}{Quark masses} \\
$m_d$(2 GeV) = 0.0047$^{+0.0005}_{-0.0004}$ GeV~\cite{Olive:2016xmw}& $m_d$(3 GeV) = 0.0043$^{+0.0005}_{-0.0004}$ GeV $^a$\\
$m_s$(2 GeV) = 0.096$^{+0.008}_{-0.004}$ GeV~\cite{Olive:2016xmw}& $m_s$(3 GeV) = 0.087$^{+0.007}_{-0.004}$ GeV $^a$\\
$m_s(\mu_W)$ = 0.057$^{+0.005}_{-0.002}$ GeV $^a$ & $m_s(\mu_t)$ = 0.054$^{+0.005}_{-0.002}$ GeV $^a$\\
$m_c(m_c)$ = 1.27(3) GeV~\cite{Olive:2016xmw} & $m_c(\mu_W)$ = 0.660(21) GeV $^a$\\
$m_c(\mu_t)$ = 0.623(20) GeV $^a$& $m_t^{\rm pole}$ = 173.21(87) GeV~\cite{Olive:2016xmw} \\
$m_t(m_t)$ = 163.427$^{+0.828}_{-0.829}$ GeV $^a$ & $m_t(\mu_W)$ = 173.276$^{+1.590}_{-1.586}$ GeV $^a$\\
\end{tabu}
\begin{tabu} to\linewidth {|X[2l] X[2l]|}
\hline
\multicolumn{2}{|c|}{CKM matrix elements~\cite{Olive:2016xmw}} \\
$\lambda$ = 0.22506(50) & $A$ = 0.811(26) \\
$\bar{\rho}$ = 0.124$^{+0.019}_{-0.018}$ & $\bar{\eta}$ = 0.356(11) \\
$\rho$ = 0.127$^{+0.019}_{-0.018}$ & $\eta$ = 0.365(11) \\
\end{tabu}
\begin{tabu} to\linewidth {|X[2l] X[2l]|}
\hline
\multicolumn{2}{|c|}{Kaon mixing parameters} \\
$m_K$ = 0.497611(13) GeV~\cite{Olive:2016xmw}& $F_K$ = 0.1562(9) GeV~\cite{Aoki:2016frl} \\
$\hat{B}_K$ = 0.7625(97)~\cite{Aoki:2016frl}& $B_1$(3 GeV) = 0.519(26)~\cite{Jang:2015sla} \\
$B_2$(3 GeV) = 0.525(23)~\cite{Jang:2015sla} & $B_3$(3 GeV) = 0.360(16)~\cite{Jang:2015sla} \\
$B_4$(3 GeV) = 0.981(62)~\cite{Jang:2015sla} & $B_5$(3 GeV) = 0.751(68)~\cite{Jang:2015sla} \\
$\eta_{cc}$ = 1.87(76)~\cite{Bobeth:2016llm} & $\eta_{tt}$ = 0.5765(65)~\cite{Bobeth:2016llm} \\
$\eta_{ct}$ = 0.496(47)~\cite{Bobeth:2016llm} &  \\
\end{tabu}
\begin{tabu} to\linewidth {|X[2l] X[2l]|}
\hline
\multicolumn{2}{|c|}{Experimental data~\cite{Olive:2016xmw}} \\
$(\Delta m_K)_{\rm exp}$ = 3.4839(59)$\times 10^{-15}$ GeV & $(|\epsilon_K|)_{\rm exp}$ = 2.228(11)$\times 10^{-3}$ \\
\hline\hline
\end{tabu}
\begin{tablenotes}
    \item[a] $^a$ \small This value is calculated with the \emph{RunDec} package~\cite{Chetyrkin:2000yt} at the two-loop level in $\alpha_s$.
\end{tablenotes}
\end{center}
\end{table}

Firstly, we collect in Tab.~\ref{table:input} the values of the relevant input parameters used throughout this paper, together with the experimental data on $\Delta m_K$ and $\epsilon_K$. For the bag parameters, we use the lattice results with $N_f=2+1$ flavors of dynamical quarks and evaluated at the renormalization scale $3$~GeV~\cite{Aoki:2016frl,Jang:2015sla}. In addition, we have used the \emph{RunDec} package~\cite{Chetyrkin:2000yt} to obtain the running coupling constant and quark masses at different scales in the two-loop approximation.

\begin{table}[t]
\begin{center}
\caption{\label{result:mSM} \small SM results for $\Delta m_K$ and $\epsilon_K$ with different corrections included and the ratios between these values and their experimental data. Here the column ``None" denotes the results obtained with the external strange-quark momentum and mass ignored and without including the LD contribution.}
\vspace{-0.5cm}
\def\arraystretch{1.5}
\tabcolsep 0.1in
\begin{tabular}{|l|c|c|c|c|}
\hline
\diagbox{Observables}{Corrections}& None & With LD & With $x_s$ & With LD and $x_s$ \\
\hline
$(\Delta m_K)_{\rm SM} (\times 10^{-15}\text{ GeV})$ & 3.109(1.258) & 3.458(1.258) & 3.321(1.258) & 3.670(1.258) \\
\hline
$\frac{(\Delta m_K)_{\rm SM}}{(\Delta m_K)_{\rm exp}}$ & 89.24\% & 99.24\% & 95.34\% & 105.34\% \\
\hline
$(|\epsilon_K|)_{\rm SM} (\times 10^{-3})$ & 2.219$^{+0.309}_{-0.294}$ & 2.086$^{+0.294}_{-0.280}$ & 2.218$^{+0.309}_{-0.294}$ & 2.085$^{+0.294}_{-0.280}$ \\
\hline
$\frac{(|\epsilon_K|)_{\rm SM}}{(|\epsilon_K|)_{\rm exp}}$ & 99.61\% & 93.63\% & 99.55\% & 93.58\% \\
\hline
\end{tabular}
\end{center}
\end{table}

With the input parameters collected in Tab.~\ref{table:input}, we can now give the numerical results for $\Delta m_K$ and $\epsilon_K$ in the SM case, which are listed in Tab.~\ref{result:mSM}. We make the following comments on the SM results:
\begin{itemize}
\item Our result for the mass difference $\Delta m_K$ without the corrections from the external strange-quark mass, $x_s$, and from the LD contribution, agrees well with that obtained in Ref.~\cite{Brod:2011ty}.

\item The corrections from $x_s$ to $\Delta m_K$ and $\epsilon_K$ are $6.83\%$ and $-0.06\%$, respectively. Note that the correction to $\Delta m_K$ is at the same order as that obtained in Ref.~\cite{Urban:1997cm}. Moreover, the LD contributions to $\Delta m_K$ and $\epsilon_K$ are $11.20\%$ and $-6\%$, respectively.

\item As the $x_s$ correction can be precisely calculated, we consider it both to $\Delta m_K$ and to $\epsilon_K$; especially, this correction is not too small for $\Delta m_K$. In addition, we include the LD contributions to $\epsilon_K$ but not to $\Delta m_K$, because the structure of LD contribution to $\Delta m_K$ is still not well understood~\cite{Buras:2010pza}.
\end{itemize}

\subsection{Results in the 2HDMs with MFV}

As can be seen clearly from Tab.~\ref{result:mSM}, there is no significant deviation between the SM predictions and the experimental data for $\Delta m_K$ and $\epsilon_K$, especially for the latter. Therefore, these two observables are expected to put strong constraints on the parameter spaces of the type-III and type-C 2HDMs, which are both featured by the three parameters, the two Yukawa couplings $A_{u,d}$ and the charged-Higgs mass $m_{H^\pm}$, in this paper. In the case of complex couplings, we can further choose $|A_u|$ and $A_uA_d^\ast=|A_uA_d^\ast|e^{i\theta}$ as the independent variables, with $\theta$ being the relative phase between $A_u$ and $A_d^\ast$.

The relevant model parameters are also constrained by the other processes. For the parameter $|A_u|$, an upper bound can be obtained from the $Z\to b\bar b$ decay~\cite{Degrassi:2010ne}, while the parameter $A_d$ is much less constrained phenomenologically~\cite{Degrassi:2010ne,Li:2013vlx}. However, the perturbativity of the theory requires that these couplings cannot be too large. As for the charged-Higgs mass, the lower bound $m_{H^\pm}>78.6$ GeV ($95\%$ CL) has been set by the LEP experiment~\cite{Abbiendi:2013hk}, which is obtained under the assumption that $H^\pm$ decays mainly into fermions without any specific Yukawa structure. In addition, direct searches for $H^\pm$ are also performed by the Tevatron~\cite{Gutierrez:2010zz}, ATLAS~\cite{Aad:2015nfa} and CMS~\cite{Khachatryan:2015qxa} experiments, among which most constraints depend strongly on the underling Yukawa structures. Recently, by comparing the cross-sections for the dijet, top-pair, dijet-pair, $t\bar t b\bar t$ and $b\bar b b\bar b$ production at the LHC with the strongest available experimental limits from ATLAS or CMS at $8$ or $13$~TeV, Hayreter and Valencia~\cite{Hayreter:2017wra} have extracted constraints on the parameter space of the Manohar-Wise model~\cite{Manohar:2006ga}, which is equivalent to the type-C 2HDM discussed here. Interestingly, they found that masses below $1$~TeV have not been excluded for color-octet scalars as is often claimed in the literature. For a variety of well-motivated 2HDMs, the authors in Ref.~\cite{Arbey:2017gmh} found that charged-Higgs bosons as light as $75$~GeV can still be compatible with all the results from direct charged and neutral Higgs boson searches at LEP and the LHC, as well as the most recent constraints from flavor physics, although this implies severely suppressed charged-Higgs couplings to all fermions. Thus, based on the above observations, we generate randomly numerical points for the model parameters as~\cite{Chang:2015rva}
\begin{equation}\label{eqn:random}
|A_u|\in [0,3], \qquad |A_d|\in [0,500], \qquad \theta\in [-\pi,\pi], \qquad m_{H^\pm}=100,~250,~500~\text{GeV}.
\end{equation}

Taking $m_{H^\pm}=500$~GeV as a benchmark, we firstly explore the dependence of each Wilson coefficient evaluated at the matching scale $\mu=m_{H^\pm}$ or approximately at $\mu=m_t$ on the other model parameters,
\begin{align}
C^{VLL}_{III}\times 10^9&=(41.28-42.15i)|A_u|^2+(6.97-7.19i)|A_u|^4+10^{-4}\cdot(5.33-0.05i) A_uA_d^*,\\[0.2cm]
C^{SLL}_{III}\times 10^{15}&=(3.28-3.29i)|A_u|^2-(0.09-0.09i)|A_u|^4-(308.73-25.25i) A_uA_d^*\nonumber\\
&+(0.45-0.46i)|A_u|^2A_uA_d^*-(0.45-0.46i)(A_uA_d^*)^2,\\[0.2cm]
C^{VLL}_{C}\times 10^9&=(13.76-14.05i)|A_u|^2+(4.26-4.39i)|A_u|^4+10^{-4}\cdot(1.78-0.02i) A_uA_d^*,\\[0.2cm]
C^{SLL}_{C}\times 10^{15}&=-(1.37-1.37i)|A_u|^2+(0.02-0.02i)|A_u|^4+(128.64-10.52i) A_uA_d^*\nonumber\\
&\hspace{0.35cm}-(0.12-0.12i)|A_u|^2A_uA_d^*+(0.12-0.12i)(A_uA_d^*)^2,\\[0.2cm]
C^{TLL}_{C}\times 10^{16}&=(2.05-2.06i)|A_u|^2-(0.06-0.06i)|A_u|^4-(192.96-15.78i) A_uA_d^*\nonumber\\
&+(0.32-0.33i)|A_u|^2A_uA_d^*-(0.32-0.33i)(A_uA_d^*)^2.
\end{align}
From the above numerical results, we can make the following observations:
\begin{itemize}
\item The dominant contribution to the effective Hamiltonian given by Eq.~\eqref{eqn:Hamiltonian} comes from the operator $Q^{VLL}$ in both the type-III and the type-C 2HDM, due to the $x_s$ suppression in $Q^{SLL}$ and $Q^{TLL}$. Furthermore, the coefficient of the $A_uA_d^\ast$ term in $Q^{VLL}$ is quite small, being of order $\mathcal O(10^{-4})$ compared to that of the $|A_u|$ terms.

\item Due to the color factor, the Wilson coefficient $C^{VLL}$ in the type-C is a little bit smaller than that in the type-III 2HDM, and the sign of $C^{SLL}$ in the type-C is also flipped relative to that in the type-III 2HDM.

\item There exists an extra operator $Q^{TLL}$ in the type-C 2HDM, and its Wilson coefficient $C^{TLL}$ differs from that of $Q^{SLL}$ in sign.
\end{itemize}

\begin{figure}[t]
\begin{center}
  \subfigure[]{\includegraphics[width=3.3in]{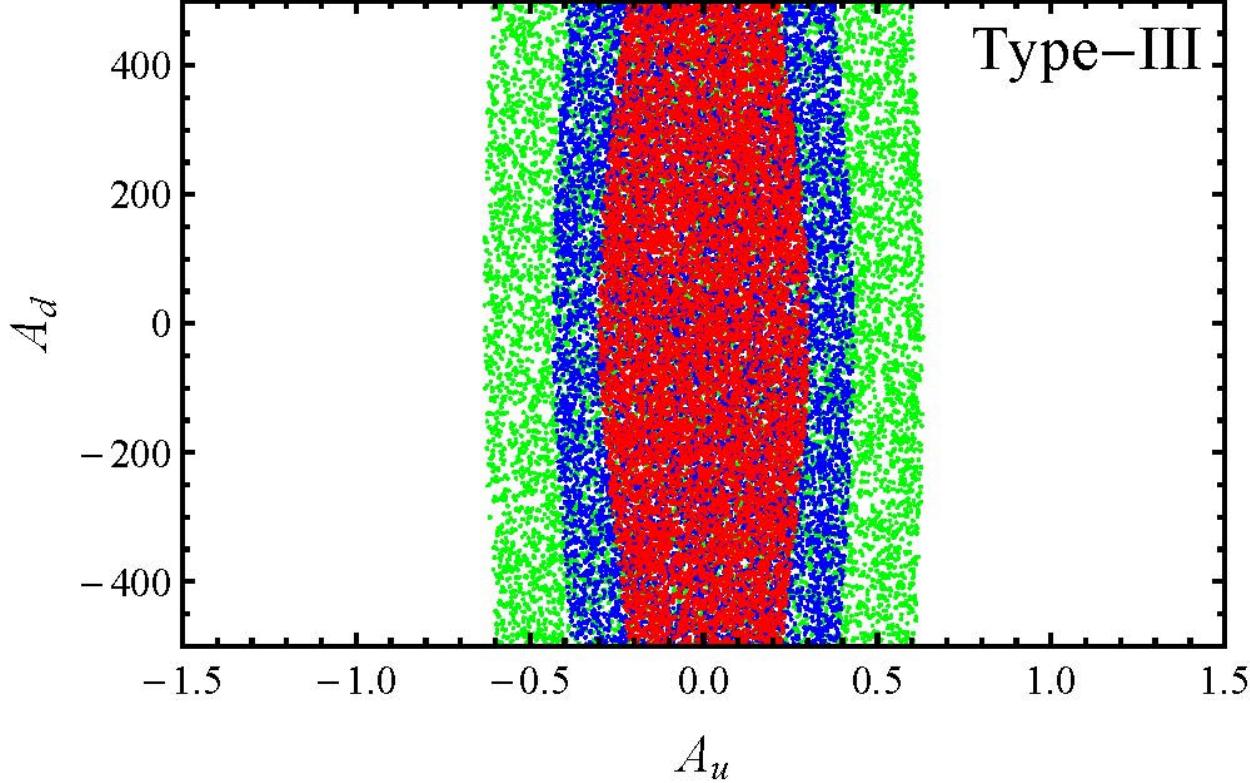}}~
  \subfigure[]{\includegraphics[width=3.3in]{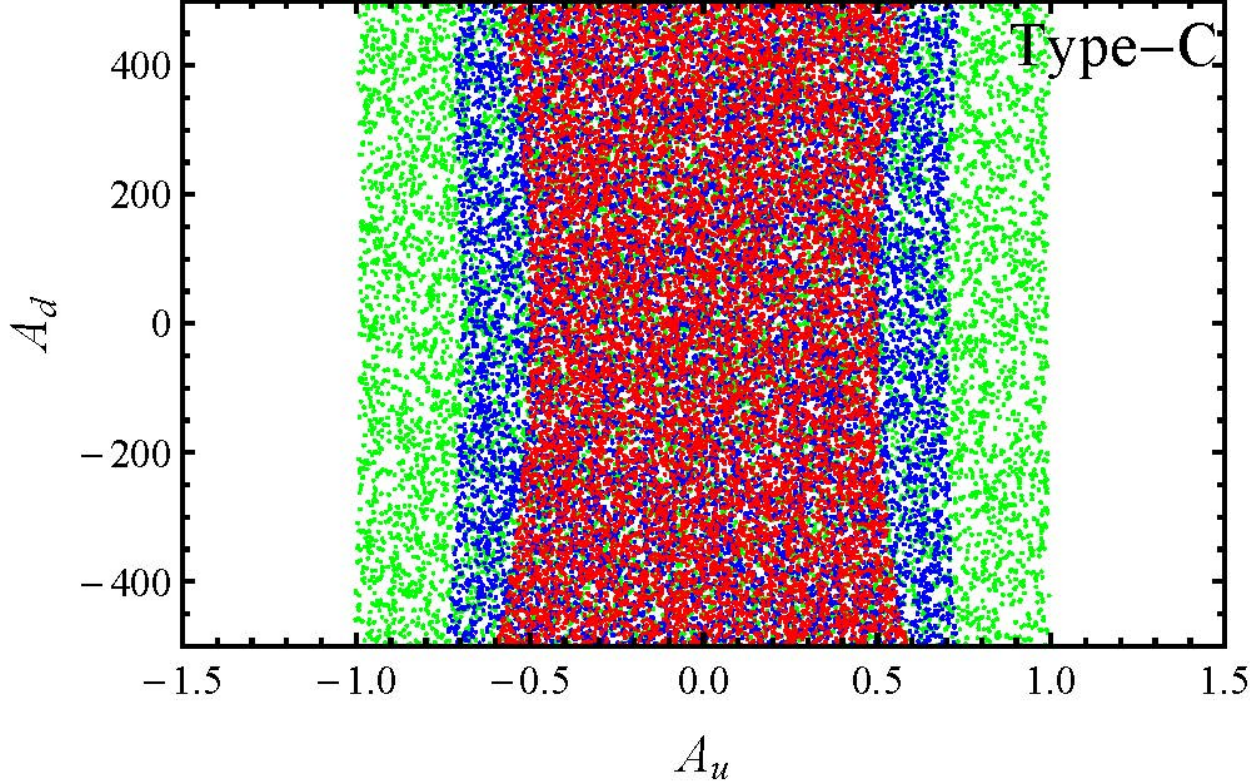}}
   \caption{\label{real result} \small Allowed parameter spaces for $A_u$ and $A_d$ in the case of real coupling for the type-III and type-C 2HDMs, under the combined constraint from $\Delta m_K$ and $\epsilon_K$. The red, blue and green regions are obtained with $m_{H^\pm}$ fixed at 100, 250 and 500 GeV, respectively.}
\end{center}
\end{figure}

\begin{figure}[t]
\begin{center}
  \subfigure[]{\includegraphics[width=3.15in]{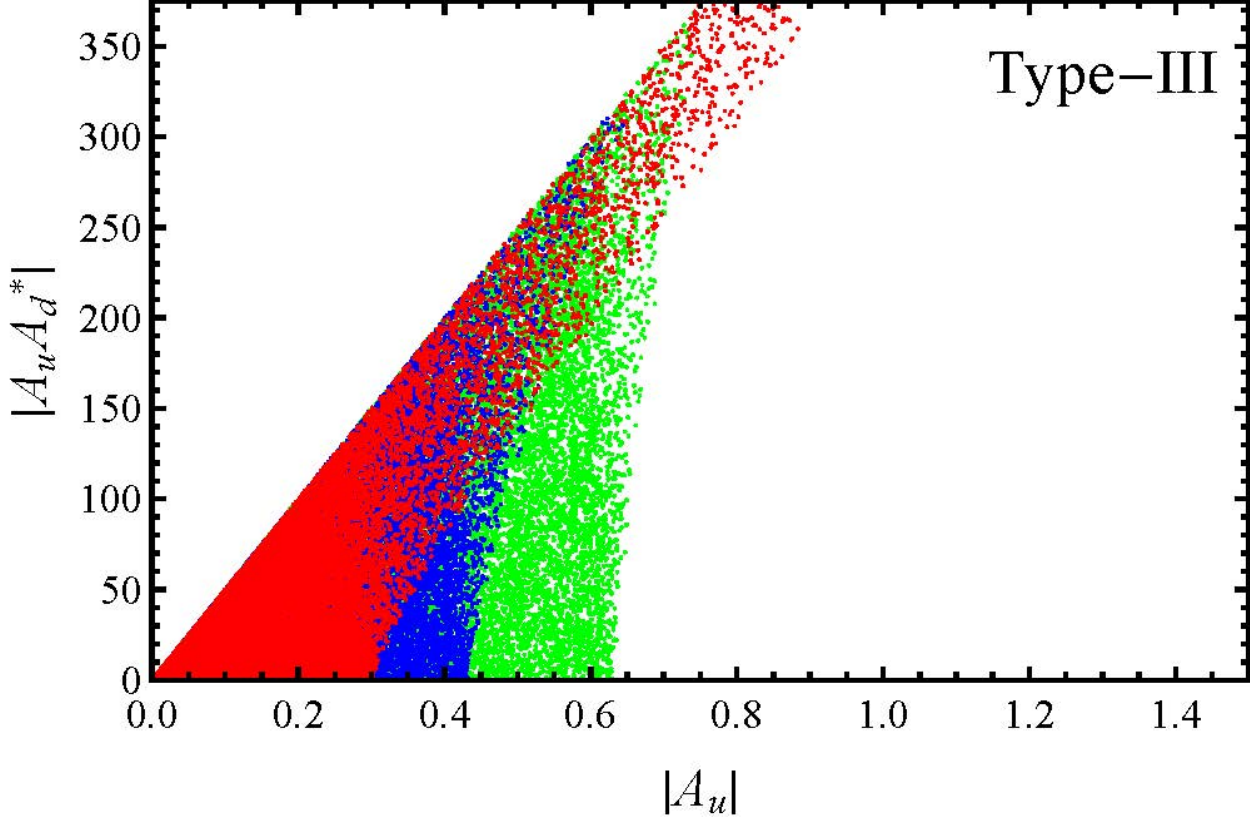}}
  \hspace{0.2in}
  \subfigure[]{\includegraphics[width=3.15in]{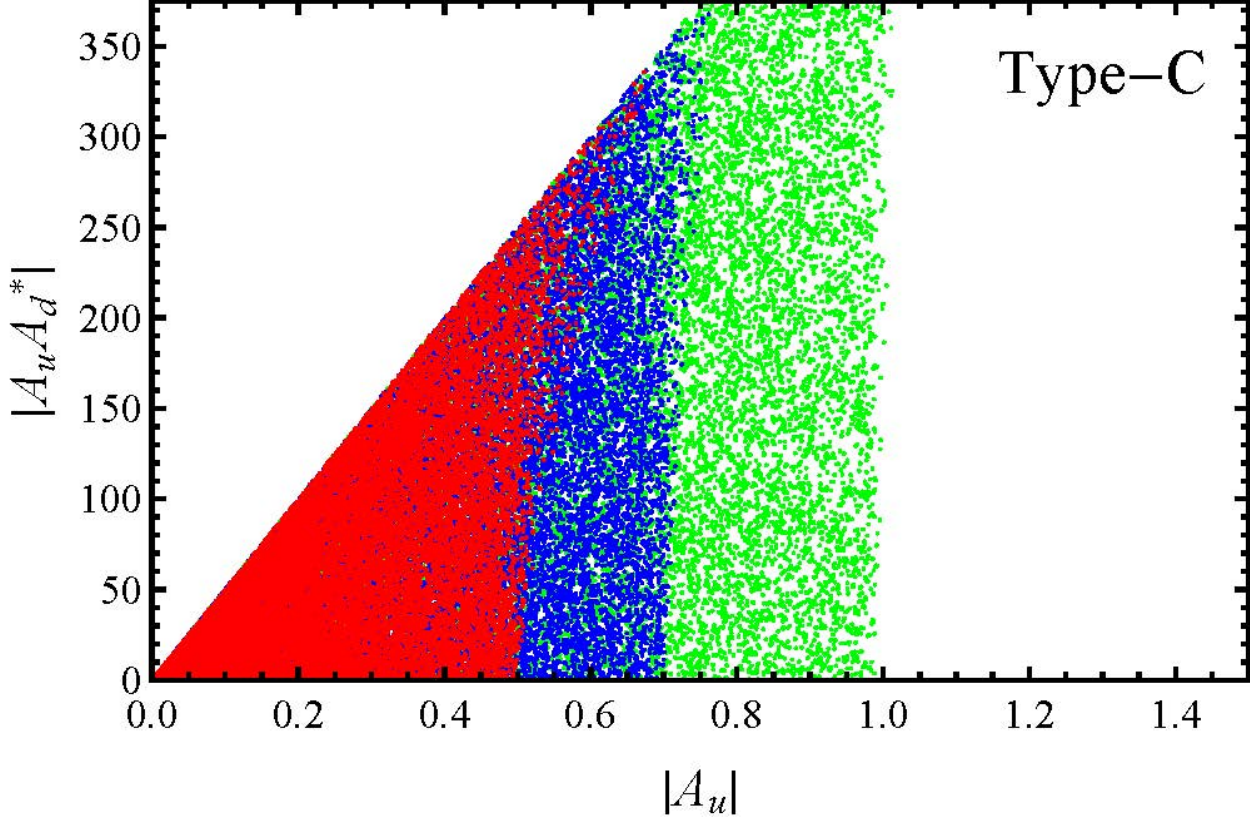}}
  \newline
  \subfigure[]{\includegraphics[width=3.15in]{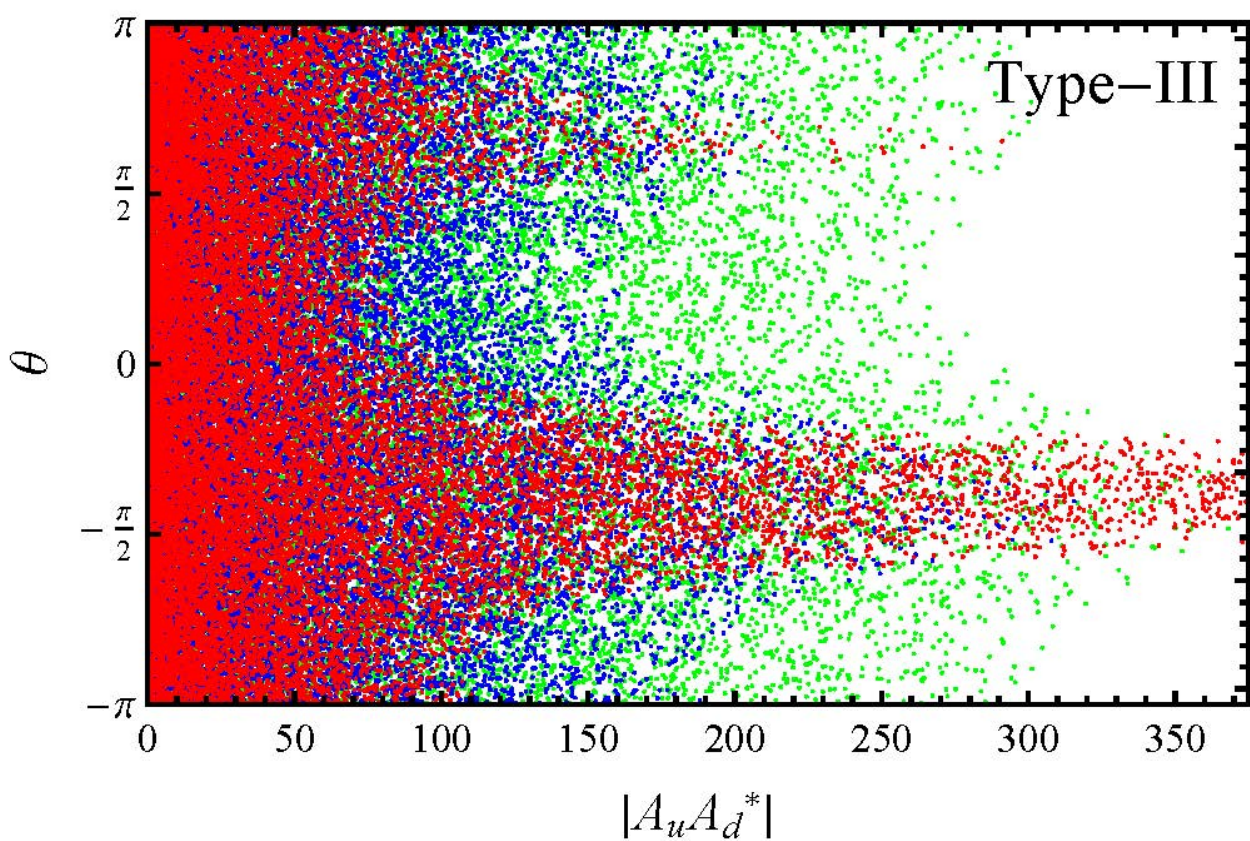}}
  \hspace{0.2in}
  \subfigure[]{\includegraphics[width=3.15in]{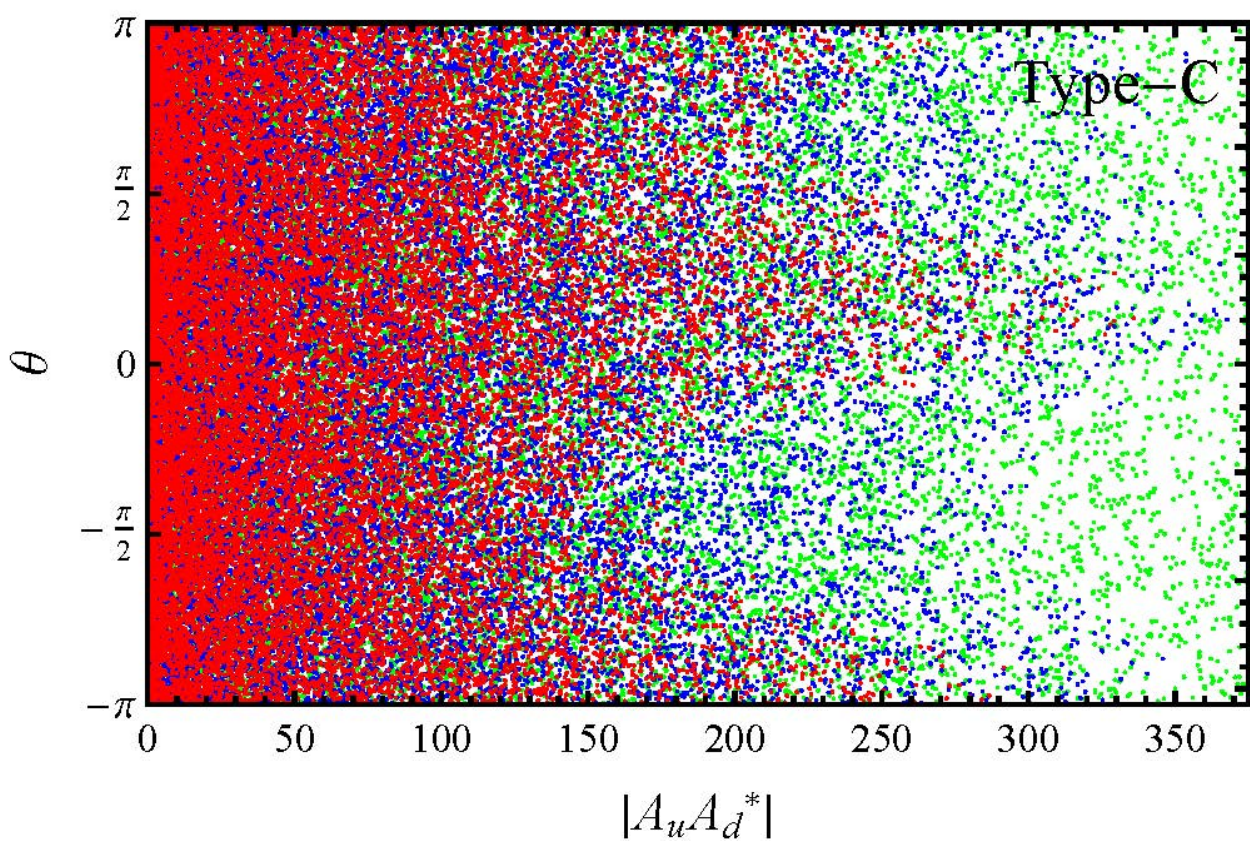}}
   \caption{\label{complex result} \small Allowed parameter spaces for $|A_u|$, $|A_uA_d^\ast|$ and $\theta$ in the case of complex coupling for the type-III and type-C 2HDMs, under the combined constraint from $\Delta m_K$ and $\epsilon_K$. The exclusion region in $|A_uA_d^\ast|$ comes from the constraint on $|A_d|$. The other captions are the same as in Fig.~\ref{real result}.}
\end{center}
\end{figure}

From the current experimental data on $\Delta m_K$ and $\epsilon_K$, one can constrain the model parameters and even distinguish the two scenarios of 2HDM with MFV. To get the plots for the allowed parameter spaces, we do as follows:
\begin{enumerate}
\item We scan the Yukawa coupling parameters $A_u$ and $A_d$ (also the relative phase $\theta$ for complex couplings) randomly within the ranges given by Eq.~\eqref{eqn:random}, with $m_{H^\pm}$ fixed at 100, 250 and 500 GeV, respectively.

\item With each set of values for the model parameters, we give the theoretical prediction for $\Delta m_K$ and $\epsilon_K$, together with the corresponding uncertainty resulted from the input parameters listed in Tab.~\ref{table:input}. The method of calculating the theoretical uncertainty is the same as in Ref.~\cite{Chang:2015rva}.

\item We select the points which lead to the theoretical predictions overlapping with the $2\sigma$ range of the experimental data.
\end{enumerate}
The final allowed spaces for the model parameters are shown in Fig.~\ref{real result} for the real coupling and in Fig.~\ref{complex result} for the complex coupling case, respectively.

From Fig.~\ref{real result}, we can make the following observations for the real coupling:
\begin{itemize}
\item In the type-III model, as shown in Fig.~\ref{real result}~(a), the parameter $A_u$ is severely constrained due to the good agreement between the SM predictions and the experimental data, especially for $\epsilon_K$; for example, the limit $|A_u|<0.7$ is more stringent compared to that obtained in Ref.~\cite{Chang:2015rva} with $m_{H^\pm}=500$~GeV. However, there is almost no constraints on $A_d$ because of the smallness of the coefficient involving $A_d$, as mentioned earlier.

\item In the type-C model, we also get strong constraint for $A_u$, but being looser than that in the type-III case, with the maximum value $|A_u|\approx 1$. The wider allowed range in the type-C model comes from the additional color factor. Similar to that observed in the type-III model, there is also almost no constraint on $A_d$.

\item The patterns of the allowed parameter spaces of these two models are different, looking like ``convex lens" for the type-III, while like ``concave lens" for the type-C model. This means that the allowed range for $|A_u|$ is smaller with larger $|A_d|$ for the type-III model, while the allowed range for $|A_u|$ in the type-C model can be larger with greater $|A_d|$.  The reason is that the dominant contribution to the two observables $\Delta m_K$ and $\epsilon_K$ comes from the operator $Q^{VLL}$, the Wilson coefficient of which in the type-C is smaller than that in the type-III model. Moreover, the cancellation between the Wilson coefficients $Q^{SLL}$ and $Q^{TLL}$ in the type-C model also reduces their contribution to the observables.
\end{itemize}
For the complex coupling, on the other hand, the results shown in Fig.~\ref{complex result} imply that
\begin{itemize}
\item In the type-III model, there exist a strong correlation between $|A_u|$ and $|A_uA_d^\ast|$, especially when $m_{H^\pm}=100$~GeV, as shown in Fig.~\ref{complex result}~(a). It is also found from Fig.~\ref{complex result}~(c) that the large values of $|A_uA_d^\ast|$ are allowed at $\theta\approx\pm\pi/2$, due to the cancellation between the complex terms.

\item In the type-C model, as shown in Figs.~\ref{complex result}~(b) and \ref{complex result}~(d), similar observations can also be made, except for the fact that the constraints on the couplings are now a little bit looser than that in the type-III model. What makes different from the type-III model is that larger values of $|A_uA_d^\ast|$ in the $|A_uA_d^\ast|-\theta$ plane occur around $\theta\approx0$ and $\pm\pi$, which are resulted from the cancellation between the complex terms.
\end{itemize}

From the above discussions, one can conclude that, although the type-III and the type-C model present some significantly different behaviors under the experimental constraints from $K^0-\bar K^0$ mixing, it is still hard to distinguish them from each other, especially for the real coupling case or for small $|A_d|$. This is due to the significant uncertainties of both the theoretical predictions and the experimental data. Therefore, more refined theoretical and experimental efforts are needed for a much clearer phenomenological picture.

\section{Conclusion}
\label{sec:conclusion}

In this paper, we have performed a complete one-loop computation of the box diagrams for the $K^0-\bar K^0$ mixing, both within the SM and in the type-III and type-C 2HDMs. It is noted that, in order to get a gauge-independent result, the external strange-quark momentum and mass should be taken into account, which has been kept up to the second order.

Combining the latest experimental data on the $K^0-\bar K^0$ mixing, we then performed a detailed phenomenological analysis of the charged-Higgs effects on this process. Our main conclusions can be summarized as follows:
\begin{itemize}
\item The operator $Q^{TLL}$ appears already at the matching scale in the type-C model, while its appearance in the type-III model is induced by the RG evolution effect from the high- down to the low-energy scale.

\item We get strong constraint on the Yukawa coupling parameter $|A_u|$ in both the real and the complex coupling case, being even stronger than that obtained in Ref.~\cite{Chang:2015rva}, while there is almost no constraint on the other Yukawa coupling $|A_d|$.

\item The allowed parameter spaces for $A_u$ and $A_d$ in the case of real coupling are similar for both types of models, with a wider range in the type-C model. If we extend the $A_d$ range, however, the allowed region for $A_u$ will be smaller in the type-III and larger in the type-C model, behaving like ``convex lens" and ``concave lens", respectively.

\item In the case of complex coupling, the strong correlation between $|A_u|$ and $|A_uA_d^\ast|$ is observed, especially for $m_{H^\pm}=100$~GeV in the type-III model. The relative phase between $|A_u|$ and $|A_d|$, $\theta$, allows the large values of $|A_uA_d^\ast|$ at $\theta\approx\pm\pi/2$ in the type-III and $\theta\approx 0$ and $\pm\pi$ in the type-C model. This is due to the cancellation effect between the complex terms in the Wilson coefficient $C^{VLL}$.
\end{itemize}

Although these two types of models present some significantly different behaviors under the experimental constraints from $K^0-\bar K^0$ mixing, it is still hard to distinguish one from the other, especially for the real coupling case or for small $|A_d|$. We need more refined theoretical and experimental efforts for a much clearer phenomenological picture.

\section*{Acknowledgements}

The work is supported by the National Natural Science Foundation of China (NSFC) under contract Nos. 11675061 and 11435003. XZ is supported by the CCNU-QLPL Innovation Fund (QLPL2015P01).

\begin{appendix}

\section*{Appendix: Basic functions for $K^0-\bar{K}^0$ mixing}

In this appendix, we collect the relevant functions during the calculation of the Wilson coefficients for $K^0-\bar{K}^0$ mixing. Note that the notation $f(x)=\lim_{y \to x} f(x,y)$ and $f(x,x_H)=\lim_{y \to x} f(x,y,x_H)$ is applied to each function listed below.

The Inami-Lim function $S_0$ is given by~\cite{Inami:1980fz}
\begin{equation}{\label{S0}}
\begin{split}
S_0(x_c,x_t)&= x_c x_t\bigg[\frac
   {\left(x_c^2-8
   x_c+4\right) \ln
   \left(x_c\right)}{4
   \left(x_c-1\right){}^2
   \left(x_c-x_t\right)}-\frac{\left(x_t^2-8
   x_t+4\right) \ln
   \left(x_t\right)}{4
   \left(x_t-1\right){}^2
   \left(x_c-x_t\right)}
   -\frac{3}{4
   \left(x_c-1\right)
   \left(x_t-1\right)}
   \bigg]
\end{split}.
\end{equation}
The functions $f_i$ introduced in sect.~\ref{sec:analytic calculation} are given explicitly as
\begin{equation}
\begin{split}
f_1(x_c,x_t)&=\frac{\ln
   \left(x_t\right)}{12
   \left(x_t-1\right){}^4
   \left(x_c-x_t\right){}^3}\bigg[x_c^3
   \left(x_t^4-9 x_t^3+36
   x_t^2-42
   x_t+12\right)\\
   &+x_c^2 x_t
   \left(-3 x_t^4+22 x_t^3-87
   x_t^2+108 x_t-36\right)+x_c
   x_t^3 \left(15 x_t^2-23
   x_t+6\right)\bigg]\\
   &+\frac{\ln
   \left(x_c\right)}{12
   \left(x_c-1\right){}^4
   \left(x_t-x_c\right){}^3}\bigg[x_t^3
   \left(x_c^4-9 x_c^3+36
   x_c^2-42
   x_c+12\right)\\
   &+x_t^2 x_c
   \left(-3 x_c^4+22 x_c^3-87
   x_c^2+108 x_c-36\right)+x_t
   x_c^3 \left(15 x_c^2-23
   x_c+6\right)\bigg]\\
   &-\dfrac{1}{72 \left(x_c-1\right){}^3 \left(x_t-1\right){}^3
   \left(x_c-x_t\right){}^2}\bigg[x_c^5 \left(65 x_t^3-130 x_t^2+113 x_t-60\right)\\
&+x_c^4 \left(-118x_t^4+34 x_t^3+250 x_t^2-298 x_t+180\right)\\
&+x_c^3 \left(65x_t^5+34 x_t^4+66 x_t^3-386 x_t^2+329 x_t-180\right)\\
&-2x_c^2\left(65 x_t^5-125 x_t^4+193 x_t^3-217 x_t^2+90x_t-30\right)\\
&+x_c x_t \left(113 x_t^4-298 x_t^3+329 x_t^2-180
   x_t+24\right)-60 \left(x_t-1\right){}^3 x_t^2\bigg],
\end{split}
\end{equation}
\begin{equation}
\begin{split}
f_2(x_c,&x_t)=\frac{x_t}{36
   \left(x_c-1\right){}^3
   \left(x_t-1\right){}^3
   \left(x_c-x_t\right){}^2}
   \bigg[x_c^5 \left(5
   x_t^2-22 x_t+5\right)\\
   &+2x_c^4 \left(x_t^3-x_t^2+35
   x_t-11\right)+x_c^3 \left(5
   x_t^4-2 x_t^3-78 x_t^2-2
   x_t+5\right)\\
   &-2 x_c^2 x_t\left(11 x_t^3-35
   x_t^2+x_t-1\right)+x_c
   \left(5 x_t^4-22 x_t^3+5
   x_t^2\right)\bigg]\\
&-\frac{x_c^3 x_t \ln \left(x_c\right)}{6\left(x_c-1\right){}^4\left(x_c-x_t\right){}^3}
\bigg[3 x_c^2
   \left(x_t+1\right)-x_c
   \left(x_t^2+10
   x_t+1\right)+3 x_t
   \left(x_t+1\right)\bigg]\\
   &-\frac{x_t^3 x_c \ln \left(x_t\right)}{6\left(x_t-1\right){}^4\left(x_t-x_c\right){}^3}
\bigg[3 x_t^2
   \left(x_c+1\right)-x_t
   \left(x_c^2+10
   x_c+1\right)+3 x_c
   \left(x_c+1\right)\bigg],
\end{split}
\end{equation}
\begin{align}
f_3(x_c,&x_t)=-\frac{1}{36
   \left(x_c-1\right){}^3
   \left(x_t-1\right){}^3
   \left(x_c-x_t\right){}^2}\bigg[x_c^5 \left(10 x_t^3-25
   x_t^2+8 x_t-5\right)\nonumber\\
   &+x_c^4\left(-20 x_t^4+25 x_t^3+49
   x_t^2-21x_t+15\right)\nonumber\\
   &+x_c^3\left(10 x_t^5+25 x_t^4-102x_t^3+2 x_t^2+8
   x_t-15\right)\nonumber\\
   &+x_c^2\left(-25 x_t^5+49 x_t^4+2x_t^3+26 x_t^2-9
   x_t+5\right)\nonumber\\
   &+x_c \left(8x_t^5-21 x_t^4+8 x_t^3-9x_t^2+2 x_t\right)-5
   \left(x_t-1\right){}^3
   x_t^2\bigg]\nonumber\\
&-\frac{\ln \left(x_t\right)}{6
   \left(x_t-1\right){}^4
   \left(x_c-x_t\right){}^3}\bigg[x_c^3 \left(3
   x_t-1\right)-x_c^2
   \left(x_t^3+6 x_t^2-3
   x_t\right)+x_c \left(3
   x_t^4-x_t^3\right)\bigg]\nonumber\\
   &-\frac{\ln \left(x_c\right)}{6
   \left(x_c-1\right){}^4
   \left(x_t-x_c\right){}^3}\bigg[x_t^3 \left(3
   x_c-1\right)-x_t^2
   \left(x_c^3+6 x_c^2-3
   x_c\right)+x_t \left(3
   x_c^4-x_c^3\right)\bigg],
\end{align}
\begin{align}
f_4(x_c,&x_t,x_H)=\frac{\left(x_c-4\right) x_c^2
   x_t\ln(x_c)}{2 \left(x_c-1\right)
   \left(x_c-x_H\right)
   \left(x_c-x_t\right)}-\frac{x_c \left(x_t-4\right)
   x_t^2\ln(x_t)}{2 \left(x_t-1\right)
   \left(x_c-x_t\right)
   \left(x_t-x_H\right)}\nonumber\\
   &-\frac{x_c \left(x_H-4\right)
   x_H x_t\ln(x_H)}{2
   \left(x_H-1\right)
   \left(x_c-x_H\right)
   \left(x_H-x_t\right)}\nonumber\\
   &+x_s\Bigg\{\frac{x_c^2 x_t\ln(x_c)}{24\left(x_c-1\right){}^3 \left(x_c-x_H\right){}^3\left(x_c-x_t\right){}^3}\bigg[-3 x_c^5 \left(x_H+2 x_t-7\right)\nonumber\\
&+x_c^4
   \bigg(4 x_H \left(4 x_t-3\right)+x_H^2+2 x_t^2+16
   x_t-23\bigg)\nonumber\\
   &-x_c^3 \bigg(x_H^2 \left(6 x_t-7\right)+x_H
   \left(5 x_t^2+76 x_t+5\right)+5 x_t^2+42 x_t-12\bigg)\nonumber\\
   &+x_c^2
   \bigg(x_H^2 \left(x_t^2+28 x_t+2\right)+8 x_H x_t \left(3
   x_t+20\right)+x_t \left(13 x_t+12\right)\bigg)\nonumber\\
   &-3 x_c x_H x_t
   \bigg(x_H \left(x_t+22\right)+17 x_t+20\bigg)+12 x_H x_t
   \bigg(x_H \left(x_t+2\right)+x_t\bigg)\bigg]\nonumber\\
   &-\frac{x_c x_t^2 \ln(x_t)}{24
   \left(x_t-1\right){}^3 \left(x_c-x_t\right){}^3
   \left(x_t-x_H\right){}^3}\bigg[x_c^2 \bigg(x_H^2 \left(x_t^2-3
   x_t+12\right)\nonumber\\
   &+x_H \left(-5 x_t^3+24 x_t^2-51
   x_t+12\right)+x_t^2 \left(2 x_t^2-5 x_t+13\right)\bigg)\nonumber\\
   &-2 x_c
   \bigg(x_H^2 \left(3 x_t^3-14 x_t^2+33 x_t-12\right)+2 x_H x_t
   \left(-4 x_t^3+19 x_t^2-40 x_t+15\right)\nonumber\\
   &+x_t^2 \left(3 x_t^3-8
   x_t^2+21 x_t-6\right)\bigg)+x_t^2 \bigg(x_H^2 \left(x_t^2+7
   x_t+2\right)\nonumber\\
   &-x_H x_t \left(3 x_t^2+12 x_t+5\right)+x_t
   \left(21 x_t^2-23 x_t+12\right)\bigg)\bigg]\nonumber\\
   &-\frac{x_c x_H x_t \ln(x_H)}{24 \left(x_H-1\right){}^3
   \left(x_c-x_H\right){}^3 \left(x_H-x_t\right){}^3}\bigg[x_c^2 \bigg(x_H^3 \left(6-5
   x_t\right)\nonumber\\
   &+x_H^2 \left(2 x_t^2+16 x_t+13\right)
   -3 x_H x_t
   \left(2 x_t+21\right)+x_H^4+12 x_t \left(2
   x_t+1\right)\bigg)\nonumber\\
   &-x_c \bigg(x_H^4 \left(8-12
   x_t\right)+x_H^3 \left(5 x_t^2+40 x_t+41\right)-4 x_H^2
   \left(4 x_t^2+39 x_t+3\right)\nonumber\\
   &+3 x_H x_t \left(21x_t+16\right)+3 x_H^5-12 x_t^2\bigg)+x_H^2 \bigg(-3 x_H^3
   \left(x_t-6\right)\nonumber\\
   &+x_H^2 \left(x_t^2-8 x_t+2\right)+x_H x_t
   \left(6 x_t-41\right)+x_t \left(13
   x_t+12\right)\bigg)\bigg]\Bigg\},
\end{align}
\begin{equation}
\begin{split}
g_4(x_c,&x_t,x_H)=-\frac{x_c x_t }{12 \left(x_c-1\right){}^2
   \left(x_c-x_H\right){}^2 \left(x_H-1\right){}^2
   \left(x_c-x_t\right){}^2 \left(x_H-x_t\right){}^2
   \left(x_t-1\right){}^2}\bigg[\\
   &x_c^5\bigg(\left(x_t^2+4\right) x_H^3+\left(-13
   x_t^2+15 x_t-17\right) x_H^2+\left(4 x_t^3-4 x_t^2+12
   x_t+3\right) x_H\\
   &+x_t \left(6 x_t^2-14 x_t+3\right)\bigg)
   -\bigg(\left(3 x_t^2-2 x_t+9\right) x_H^4-\left(2
   x_t^3+11 x_t^2-22 x_t+24\right) x_H^3\\
   &+\left(2 x_t^4-16
   x_t^3-11 x_t^2+31 x_t-21\right) x_H^2+\left(4 x_t^4+11
   x_t^3+x_t^2+13 x_t+6\right) x_H\\
   &+x_t \left(14 x_t^3-23 x_t^2-12
   x_t+6\right)\bigg) x_c^4+x_c^3\bigg(\left(x_t^2+4\right)
   x_H^5+\left(23 x_t^2-25 x_t+17\right) x_H^4\\
   &+2 \left(x_t^4-32
   x_t^3+14 x_t^2+22 x_t-35\right) x_H^3+8 \left(2 x_t^4+x_t^3-4
   x_t^2+4 x_t+2\right) x_H^2\\
   &+\left(4 x_t^5-11 x_t^4+8 x_t^3+35
   x_t^2-24 x_t+3\right) x_H+x_t \left(6 x_t^4+23 x_t^3-72
   x_t^2+25 x_t+3\right)\bigg)\\
   &+\bigg(\left(x_t^3-18
   x_t^2+17 x_t-15\right) x_H^5+\left(-3 x_t^4+23 x_t^3-22
   x_t^2-2 x_t+19\right) x_H^4\\
   &+\left(x_t^5+11 x_t^4+28 x_t^3-16
   x_t^2-5 x_t+21\right) x_H^3\\
   &-\left(13 x_t^5-11 x_t^4+32
   x_t^3+16 x_t^2-5 x_t+15\right) x_H^2\\
   &-x_t \left(4
   x_t^4+x_t^3-35 x_t^2+30 x_t-15\right) x_H+x_t^2 \left(-14
   x_t^3+12 x_t^2+25 x_t-18\right)\bigg) x_c^2\\
   &+\bigg(\left(17
   x_t^2-8 x_t+6\right) x_H^5+\left(2 x_t^4-25 x_t^3-2 x_t^2+2
   x_t-12\right) x_H^4\\
   &+\left(-22 x_t^4+44 x_t^3-5 x_t^2-8
   x_t+6\right) x_H^3+x_t \left(15 x_t^4-31 x_t^3+32 x_t^2+5
   x_t-6\right) x_H^2\\
   &+x_t^2 \left(12 x_t^3-13 x_t^2-24
   x_t+15\right) x_H+3 \left(x_t-1\right){}^2 x_t^3\bigg)
   x_c\\
   &+x_H x_t \bigg(\left(4 x_t^2-15 x_t+6\right) x_H^4+\left(-9
   x_t^3+17 x_t^2+19 x_t-12\right) x_H^3\\
   &+\left(4 x_t^4+24
   x_t^3-70 x_t^2+21 x_t+6\right) x_H^2+x_t \left(-17 x_t^3+21
   x_t^2+16 x_t-15\right) x_H\\
   &+3 \left(x_t-1\right){}^2
   x_t^2\bigg)\bigg],
\end{split}
\end{equation}
\begin{align}
f_5(x_c,&x_t,x_H)=-\frac{x_c\ln(x_c)}{4 \left(x_c-1\right){}^2 \left(x_c-x_H\right){}^2
   \left(x_c-x_t\right)}\bigg[-x_c^2 \left(6 x_H+11 x_t+6\right)\nonumber\\
   &+2 x_c
   \bigg(x_H \left(6 x_t+3\right)+5 x_t\bigg)+6 x_c^3-11 x_H
   x_t\bigg]+\frac{x_c x_t\ln(x_t)}{4 \left(x_t-1\right){}^2 \left(x_c-x_t\right)
   \left(x_H-x_t\right){}^2}\nonumber\\
   &\bigg[x_H \left(6 x_t-5\right)+\left(4-5 x_t\right)
   x_t\bigg]+\frac{x_c \ln(x_H)}{4 \left(x_H-1\right){}^2
   \left(x_c-x_H\right){}^2 \left(x_H-x_t\right){}^2}\bigg[\nonumber\\
   &x_c \bigg(-x_H^2 \left(19 x_t+6\right)+x_H x_t
   \left(12 x_t+17\right)+6 x_H^3-10 x_t^2\bigg)\nonumber\\
   &+x_H \bigg(x_H^2
   \left(20 x_t+6\right)-x_H x_t \left(13 x_t+18\right)-6
   x_H^3+11 x_t^2\bigg)\bigg]\nonumber\\
   &-\frac{x_c x_t}{4 \left(x_c-1\right)
   \left(x_H-1\right) \left(x_t-1\right) \left(x_c-x_H\right)
   \left(x_H-x_t\right)}\bigg[\nonumber\\
   &x_c \left(x_H-2 x_t+1\right)+x_H
   x_t-x_H^2+x_t-1\bigg],
\end{align}
\begin{equation}
\begin{split}
f_6(x_c,&x_t,x_H)=\frac{x_c^3 x_t\ln(x_c)}{4
   \left(x_c-x_H\right){}^2
   \left(x_c-x_t\right)}-\frac{x_c x_t^3\ln(x_t)}{4
   \left(x_c-x_t\right)
   \left(x_H-x_t\right){}^2}\\
   &-\frac{x_c x_H x_t\ln(x_H)}{4
   \left(x_c-x_H\right){}^2
   \left(x_H-x_t\right){}^2}\bigg[x_c
   \left(x_H-2 x_t\right)+x_H
   x_t\bigg]-\frac{x_c x_H x_t}{4
   \left(x_c-x_H\right)
   \left(x_H-x_t\right)}\\
   &+x_s\bigg\{\frac{x_c^3 x_t\ln(x_c)
   }{12
   \left(x_c-x_H\right){}^4
   \left(x_c-x_t\right){}^3}\bigg[3 x_c^2
   \left(x_H+x_t\right)-x_c
   \left(10 x_H
   x_t+x_H^2+x_t^2\right)\\
   &+3
   x_H x_t
   \left(x_H+x_t\right)\bigg]+\frac{x_c x_t^3\ln(x_t)}{12
   \left(x_c-x_t\right){}^3
   \left(x_H-x_t\right){}^4}\bigg[x_c^2
   \left(x_t-3 x_H\right)\\
   &+x_c
   \left(10 x_H x_t-3 x_H^2-3
   x_t^2\right)+x_H x_t
   \left(x_H-3
   x_t\right)\bigg]+\frac{x_c x_t\ln(x_H)}{12
   \left(x_c-x_H\right){}^4
   \left(x_H-x_t\right){}^4}\bigg[\\
   &-3 x_c
   x_H^4 \left(x_H-3
   x_t\right)+3 x_c^2 x_H
   x_t^2 \left(x_t-3
   x_H\right)+x_c^3 x_t^2
   \left(3
   x_H-x_t\right)+x_H^5
   \left(x_H-3
   x_t\right)\bigg]\\
   &+\frac{x_c x_t}{72
   \left(x_c-x_H\right){}^3
   \left(x_c-x_t\right){}^2
   \left(x_H-x_t\right){}^3}\bigg[x_c^4
   \left(-22 x_H x_t+5 x_H^2+5
   x_t^2\right)\\
   &+x_c^3 \left(70
   x_H^2 x_t-2 x_H x_t^2-22
   x_H^3+2 x_t^3\right)\\
   &+x_c^2
   \left(-2 x_H^3 x_t-78 x_H^2
   x_t^2-2 x_H x_t^3+5 x_H^4+5
   x_t^4\right)\\
   &+2 x_c x_H x_t
   \left(-x_H^2 x_t+35 x_H
   x_t^2+x_H^3-11
   x_t^3\right)+x_H^2 x_t^2
   \left(-22 x_H x_t+5 x_H^2+5
   x_t^2\right)\bigg]\bigg\},
\end{split}
\end{equation}
\begin{align}
f_7(x_c,&x_t,x_H)=-\frac{x_c x_t}
   {12 \left(x_c-1\right){}^2
   \left(x_c-x_H\right){}^2
   \left(x_H-1\right){}^2
   \left(x_c-x_t\right){}^2
   \left(x_H-x_t\right){}^2
   \left(x_t-1\right){}^2}\bigg[\nonumber\\
   &\bigg(\left(x_t^2-3
   x_t+4\right) x_H^3+\left(-3
   x_t^3+5 x_t^2-8\right)
   x_H^2+x_t \left(4 x_t^2-13
   x_t+15\right) x_H\nonumber\\
   &+x_t^2
   \left(3 x_t-5\right)\bigg)
   x_c^5+\bigg(\left(3 x_t^2-4
   x_t-3\right) x_H^4+x_t
   \left(-7 x_t^2+5
   x_t+8\right) x_H^3\nonumber\\
   &+\left(10
   x_t^4+x_t^3-16 x_t^2-4
   x_t+15\right) x_H^2+x_t
   \left(-16 x_t^3+19 x_t^2+11
   x_t-28\right) x_H\nonumber\\
   &-x_t^2
   \left(2
   x_t^2+x_t-9\right)\bigg)
   x_c^4+\bigg(\left(-2
   x_t^2+3 x_t+1\right)
   x_H^5+\left(6 x_t^3-10
   x_t^2+5 x_t+5\right)
   x_H^4\nonumber\\
   &+\left(-7 x_t^4+8
   x_t^3+4 x_t^2-16
   x_t-13\right)
   x_H^3+\left(-3
   x_t^5+x_t^4-16 x_t^3+28
   x_t^2+11 x_t-5\right)
   x_H^2\nonumber\\
   &+x_t \left(4 x_t^4+19
   x_t^3-40 x_t^2+14
   x_t+9\right) x_H+x_t^2
   \left(3 x_t^3-x_t^2-6
   x_t-2\right)\bigg)
   x_c^3\nonumber\\
   &+\bigg(\left(-2
   x_t^3+6 x_t^2-7
   x_t-3\right) x_H^5+\left(3
   x_t^4-10 x_t^3+8
   x_t^2+x_t+4\right)
   x_H^4\nonumber\\
   &+\left(x_t^5+5 x_t^4+4
   x_t^3-16 x_t^2+19
   x_t+3\right) x_H^3+x_t
   \left(5 x_t^4-16 x_t^3+28
   x_t^2-40 x_t-1\right)
   x_H^2\nonumber\\
   &+x_t^2 \left(-13
   x_t^3+11 x_t^2+14
   x_t-6\right) x_H+x_t^3
   \left(-5 x_t^2+9
   x_t-2\right)\bigg)
   x_c^2\nonumber\\
   &+x_H x_t \bigg(\left(3
   x_t^2-7 x_t+10\right)
   x_H^4+\left(-4 x_t^3+5
   x_t^2+x_t-16\right)
   x_H^3\nonumber\\
   &+\left(-3 x_t^4+8
   x_t^3-16 x_t^2+19
   x_t-2\right) x_H^2-x_t
   \left(4 x_t^2-11
   x_t+1\right) x_H\nonumber\\
   &+x_t^2
   \left(15 x_t^2-28
   x_t+9\right)\bigg)
   x_c+x_H^2 x_t^2
   \bigg(\left(x_t-3\right)
   x_H^3+\left(-3 x_t^2+5
   x_t+4\right) x_H^2\nonumber\\
   &+\left(4
   x_t^3-13 x_t+3\right)
   x_H+x_t \left(-8 x_t^2+15
   x_t-5\right)\bigg)\bigg],
\end{align}
\begin{equation}
\begin{split}
g_7(x_c,&x_t,x_H)=-\frac{x_c^2 x_t\ln(x_c)}{6
   \left(x_c-1\right){}^3
   \left(x_c-x_H\right){}^3
   \left(x_c-x_t\right){}^3}\bigg[6 x_c^6-3
   x_c^5 \left(5 x_H+5
   x_t+4\right)\\
   &+x_c^4 \bigg(5
   x_H \left(7 x_t+6\right)+8
   x_H^2+7 x_t^2+29
   x_t+5\bigg)-x_c^3 \bigg(2
   x_H^2 \left(9
   x_t+8\right)\\
   &+x_H \left(16
   x_t^2+68 x_t+13\right)+x_t
   \left(13
   x_t+12\right)\bigg)+x_c^2
   \bigg(x_H^2 \left(8
   x_t^2+35 x_t+7\right)\\
   &+x_H
   x_t \left(30
   x_t+29\right)+5
   x_t^2\bigg)-3 x_c x_H x_t
   \bigg(5 x_H
   \left(x_t+1\right)+4
   x_t\bigg)+6 x_H^2
   x_t^2\bigg]\\
   &-\frac{x_t^2 x_c\ln(x_t)}{6
   \left(x_t-1\right){}^3
   \left(x_t-x_H\right){}^3
   \left(x_t-x_c\right){}^3}\bigg[6 x_t^6-3
   x_t^5 \left(5 x_H+5
   x_c+4\right)\\
   &+x_t^4 \bigg(5
   x_H \left(7 x_c+6\right)+8
   x_H^2+7 x_c^2+29
   x_c+5\bigg)-x_t^3 \bigg(2
   x_H^2 \left(9
   x_c+8\right)\\
   &+x_H \left(16
   x_c^2+68 x_c+13\right)+x_c
   \left(13
   x_c+12\right)\bigg)+x_t^2
   \bigg(x_H^2 \left(8
   x_c^2+35 x_c+7\right)\\
   &+x_H
   x_c \left(30
   x_c+29\right)+5
   x_c^2\bigg)-3 x_t x_H x_c
   \bigg(5 x_H
   \left(x_c+1\right)+4
   x_c\bigg)+6 x_H^2
   x_c^2\bigg]\\
   &+\frac{x_c x_H x_t\ln(x_H)}{6
   \left(x_H-1\right){}^3
   \left(x_c-x_H\right){}^3
   \left(x_H-x_t\right){}^3}\bigg[x_c^2
   \bigg(-x_H^3 \left(16
   x_t+15\right)+x_H^2 \left(7
   x_t^2+29 x_t+5\right)\\
   &-3 x_H
   x_t \left(4 x_t+3\right)+8
   x_H^4+3 x_t^2\bigg)+x_c
   x_H \bigg(x_H^3 \left(36
   x_t+35\right)-x_H^2
   \left(16 x_t^2+68
   x_t+13\right)\\
   &+x_H x_t
   \left(29 x_t+24\right)-18
   x_H^4-9 x_t^2\bigg)+x_H^2
   \bigg(-18 x_H^3
   \left(x_t+1\right)+x_H^2
   \left(8 x_t^2+35
   x_t+7\right)\\
   &-x_H x_t
   \left(15 x_t+13\right)+9
   x_H^4+5
   x_t^2\bigg)\bigg],
\end{split}
\end{equation}
\begin{equation}
\begin{split}
f_8(x_c,&x_t,x_H)=\frac{x_c x_t}{4
   \left(x_c-1\right)
   \left(x_H-1\right)
   \left(x_t-1\right)
   \left(x_c-x_H\right)
   \left(x_H-x_t\right)}\bigg[\\
   &x_H
   \bigg(x_c
   \left(x_t-2\right)-2
   x_t+1\bigg)+x_c
   x_t+x_H^2\bigg]-\frac{x_c^2\ln(x_c)}{4
   \left(x_c-1\right){}^2
   \left(x_c-x_H\right){}^2
   \left(x_c-x_t\right)}\bigg[\\
   &-x_c^2
   \left(9 x_H+16
   x_t+9\right)+x_c \bigg(x_H
   \left(17 x_t+9\right)+15
   x_t\bigg)+9 x_c^3-16 x_H
   x_t\bigg]\\
   &+\frac{x_c x_t^2\ln(x_t)}{4
   \left(x_t-1\right){}^2
   \left(x_c-x_t\right)
   \left(x_H-x_t\right){}^2}\bigg[x_H
   \left(8
   x_t-7\right)+\left(6-7
   x_t\right) x_t\bigg]\\
   &+\frac{x_c x_H \ln(x_H)}{4
   \left(x_H-1\right){}^2
   \left(x_c-x_H\right){}^2
   \left(x_H-x_t\right){}^2}\bigg[\\
   &x_c
   \bigg(-x_H^2 \left(26
   x_t+9\right)+8 x_H x_t
   \left(2 x_t+3\right)+9
   x_H^3-14 x_t^2\bigg)+x_H
   \bigg(9 x_H^2 \left(3
   x_t+1\right)\\
   &-x_H x_t
   \left(17 x_t+25\right)-9
   x_H^3+15
   x_t^2\bigg)\bigg],
\end{split}
\end{equation}
\begin{align}
\begin{split}
f_9(x_c,&x_t,x_H)=-\frac{x_c^3 x_t\ln(x_c)
   }{6
   \left(x_c-x_H\right){}^4
   \left(x_c-x_t\right){}^3}\bigg[3 x_c^2
   \left(x_H+x_t\right)-x_c
   \left(10 x_H
   x_t+x_H^2+x_t^2\right)\\
   &+3
   x_H x_t
   \left(x_H+x_t\right)\bigg]-\frac{x_c x_t^3\ln(x_t)}{6
   \left(x_c-x_t\right){}^3
   \left(x_H-x_t\right){}^4}\bigg[x_c^2
   \left(x_t-3 x_H\right)\\
   &+x_c
   \left(10 x_H x_t-3 x_H^2-3
   x_t^2\right)+x_H x_t
   \left(x_H-3
   x_t\right)\bigg]+\frac{x_c x_t \ln(x_H)}{6
   \left(x_c-x_H\right){}^4
   \left(x_H-x_t\right){}^4}\bigg[\\
   &3 x_c
   x_H^4 \left(x_H-3
   x_t\right)+3 x_c^2 x_H
   x_t^2 \left(3
   x_H-x_t\right)+x_c^3 x_t^2
   \left(x_t-3
   x_H\right)+x_H^5 \left(3
   x_t-x_H\right)\bigg]\\
   &-\frac{x_c x_t}{36
   \left(x_c-x_H\right){}^3
   \left(x_c-x_t\right){}^2
   \left(x_H-x_t\right){}^3}\bigg[x_c^4
   \left(-22 x_H x_t+5 x_H^2+5
   x_t^2\right)\\
   &+x_c^3 \left(70
   x_H^2 x_t-2 x_H x_t^2-22
   x_H^3+2 x_t^3\right)\\
   &+x_c^2
   \left(-2 x_H^3 x_t-78 x_H^2
   x_t^2-2 x_H x_t^3+5 x_H^4+5
   x_t^4\right)\\
   &+2 x_c x_H x_t
   \left(-x_H^2 x_t+35 x_H
   x_t^2+x_H^3-11
   x_t^3\right)+x_H^2 x_t^2
   \left(-22 x_H x_t+5 x_H^2+5
   x_t^2\right)\bigg],
\end{split}\\[0.2cm]
\begin{split}
f_{10}(x_c,&x_t,x_H)=\frac{x_c^2
   x_t\ln(x_c)}{\left(x_c-x_H\right){}
   ^2 \left(x_c-x_t\right)}-\frac{x_c
   x_t^2\ln(x_t)}{\left(x_c-x_t\right)
   \left(x_H-x_t\right){}^2}\\
   &+\frac{x_c x_t\ln(x_H) }{\left(x_c
   -x_H\right){}^2
   \left(x_H-x_t\right){}^2}\bigg[x_c
   x_t-x_H^2\bigg]-\frac{x_c
   x_t}{\left(x_c-x_H\right)
   \left(x_H-x_t\right)},
\end{split}
\end{align}

\end{appendix}

\vspace{-1mm}
\centerline{\rule{80mm}{0.1pt}}
\vspace{2mm}

\providecommand{\href}[2]{#2}

\end{document}